\documentclass[nofootinbib,aps,11pt]{revtex4-1}
\usepackage{graphicx}
\usepackage{hyperref}
\usepackage{cancel}
\usepackage{amssymb}
\usepackage{textcomp}
\usepackage{amsmath}
\usepackage{bm}
\usepackage{times}
\usepackage{epsfig}
\usepackage{color}
\usepackage[utf8]{inputenc}
\usepackage{mathrsfs}
\usepackage{esint}
\usepackage{ulem}
\usepackage{subfigure}
\usepackage{makecell}
\usepackage{booktabs}
\usepackage{multirow}

\newcommand{\eV}{{\rm eV}}

\newcommand{\GeV}{{\rm GeV}}
\newcommand{\TeV}{{\rm TeV}}

\begin{document}
\title{\LARGE $U(1)_{B-3L_{\alpha}}$ Extended Scotogenic Models and Single-zero Textures of Neutrino Mass Matrices }
\bigskip
\author{Weijian Wang$^1$}
\email{wjnwang96@aliyun.com}
\author{Zhi-Long Han$^2$}
\email{sps\_hanzl@ujn.edu.cn}
\affiliation{
$^1$Department of Physics, North China Electric Power University, Baoding 071003, China
\\
$^2$School of Physics and Technology, University of Jinan, Jinan, Shandong 250022, China
}
\date{\today}

\begin{abstract}
In this paper, we propose a scotogenic model of neutrino masses with flavor-dependent $U(1)_{B-3L_\alpha}(\alpha=e, \mu, \tau)$ gauge symmetry, in which neutrinos are generated at
one-loop level and a fermion dark matter is naturally accommodated. In this model, three one-zero-texture structures of neutrino mass matrix, denoted as patterns A, B and C, are successfully realized by appropriate charge assignment for inert fermions and scalar fields. For each predicted textures, a detailed numerical analysis is carried out to find the allowed regions of neutrino mixing and masses. We find three scenarios (A for normal mass hierarchy, and B and C for inverted mass hierarchy) favored by the latest data of neutrino oscillation experiments and Planck 2018 limit on the sum of neutrino masses.
Other phenomenologies such as lepton flavor violation, dark matter and collider signatures are also discussed.
\end{abstract}

\maketitle

\section {Introduction}
Thanks to the overwhelming evidence from astrophysics, cosmology and neutrino oscillation experiments, it is well established that the new physics beyond Standard Model (SM) is needed to accommodate two missing spices: tiny but nonzero neutrino masses and dark matter (DM). The scotogenic model, originally proposed by by Krauss \textsl{et al}\cite{Krauss:2002px} and Ma\cite{Ma:2006km}, is one of the appealing ways to explain the above two issues simultaneously, where the small neutrino masses arise only at loop level and
DM plays a role of mediated field propagating inside the loop diagram. For a review on various interesting realizations, see Ref. \cite{Cai:2017jrq}.

In addition to the neutrino masses and DM mystery, the SM gauge symmetries cannot explain how different flavors of leptons distinguish each other and exhibit the observed mixing patterns. A promising ansatz which can potentially solves the flavor puzzle is to add an extra $U(1)$ gauge symmetry to SM and permit the three flavors of fermions transform nonuniversally under the new symmetry. On the other hand,
three right-handed(RH) neutrinos $N_R$s are usually considered as necessary ingredients of neutrino mass model. Then assuming family nonuniversal charges of $N_R$s, the resulting lepton flavor structure admits a number of anomaly-free solutions.  Along this thought of idea, the $U(1)_X$ gauge symmetry with $X$ being the linear combinations of baryon number $B$ and individual lepton numbers $L_\alpha$ is frequently considered in various contexts. Well-known examples are $B-3L_{\tau}$\cite{Ma:1997nq}, $ L_{\mu}-L_{\tau} $\cite{Ma:2001md,Choubey:2004hn, Altmannshofer:2014cfa,Heeck:2014qea}, $B+3(L_e-L_\mu-L_\tau )$\cite{Heeck:2012cd} etc.
In Ref.\cite{Araki:2012ip}, the neutrino mass matrices containing two-zero-texture\cite{Frampton:2002yf,Fritzsch:2011qv} or two-zero-cofactor structures\cite{Lavoura:2004tu} are realized by applying the $U(1)_{aB-\Sigma x_\alpha L_\alpha}$ gauge symmetry to type-I seesaw scenario. In Ref.\cite{Cebola:2013hta}, more solutions are found in the type-I and/or-III seesaw framework. It is then natural to ask if the flavor-dependent $U(1)$ gauge symmetry is compatible with the scotogenic models and leads to predictive lepton mass textures. The first attempts was made in Ref.\cite{Baek:2015mna}, where the $U(1)_{L_{\mu}-L_{\tau}}$ gauge symmetry was adopted to realize the type-C two-zero-texture structure\cite{Fritzsch:2011qv} in one-loop induced neutrino mass matrix.  Note that this model is also possible to interpret  the $R_{K^{(*)}}$ anomaly and AMS-02 positron excess \cite{Han:2019diw}.
However, the latest analysis points out that such texture structure is incompatible with the Planck 2018 bound on the sum of neutrino masses. i.e $\Sigma m_i <0.12$eV\cite{Aghanim:2018eyx}. One way out of this problem is to consider different $U(1)$ gauge symmetries. In Refs.\cite{Nomura:2018rvy,Han:2019lux}, other viable two-zero-texture structures are realized based on $U(1)_{B-2L\alpha-L_{\beta}}$ gauge symmetry.

In this paper, we apply the $U(1)_{B-3L_{\alpha}} (\alpha=e, \mu, \tau)$ gauge symmetry to the scotogenic model. It is pointed out that by pairing one of  $N_R$s with $\nu_{L\alpha}$, the $U(1)_{B-3L_{\alpha}}$ gauged symmetry is anomaly free and may predict a realistic mixing structure among three families of leptons\cite{Ma:1997nq}.
In the scotogenic model, we find that the form of Yukawa interactions is restricted by the $U(1)_{B-3L_\alpha}$ charge assignments, leading to a one-zero-texture structure of neutrino mass matrix.  For each predicted texture, an updated phenomenological analysis is carried out in view of the current neutrino oscillation and cosmological data. With all new particles around TeV scale, the scotogenic model with a $B-3L_\alpha$ gauge symmetry leads to testable phenomenologies. Thus, in addition to the neutrino issue, we discuss the phenomenologies such as lepton flavor violations (LFVs), and DM candidate and highlights of collider signatures. Without loss of generality, the $U(1)_{B-3L_\mu}$ gauge symmetry are considered as a case study.

The rest of this paper is organized  as follows. In Sec. II, we first discuss the realization of one-zero texture in the scotogenic model with $U(1)_{B-3L_\alpha}$ gauge symmetry. In Sec. III, the numerical analysis of neutrino masses and mixing is presented. In Sec. IV, other predictions such as the lepton-flavor-violation rate, dark matter and collider physics are discussed. The conclusion is given in Sec.V.

\section{Model Setup}
\begin{center}
\begin{table}
\begin{tabular}{|c||c|c|c|c|c|c|c|c|c||c|c|c|c|}
\hline\hline
\multirow{2}{*}{Group} &\multicolumn{9}{c||}{Lepton Fields} & \multicolumn{3}{c|}{Scalar Fields} \\
 \cline{2-13}
 &~$L_\alpha$~ & ~$\ell_{\alpha R}$~& ~$L_\beta$ ~ &~$\ell_{\beta R}$ ~ &~$L_\gamma$ ~&~$\ell_{\gamma R}$ ~&~$N_{R1}$ ~&~$N_{R2}$ ~&~$N_{R3}$~& ~$\Phi$ ~ & ~$\eta$~ & ~$S$~ \\ \hline
~$SU(2)_L$& ~$2$~ & ~$1$~& ~$2$ ~ &~$1$ ~ &~$2$ ~&~$1$ ~&~$1$ ~&~$1$ ~&~$1$~& ~$2$ ~ & ~$2$~ &  ~$1$
\\ \hline
~$U(1)_{Y}$& ~$-\frac{1}{2}$~ & ~$-1$~ & ~$-\frac{1}{2}$~ & ~$-1$~ & ~$-\frac{1}{2}$~ & ~$-1$~ & ~$0$~ & ~$0$~& ~$0$~ & ~$\frac{1}{2}$~ & ~$\frac{1}{2}$~& ~$0$~
\\ \hline
~$Z_2$& ~$+$~ & ~$+$~& ~$+$ ~ &~$+$ ~ &~$+$ ~&~$+$ ~&~$-$ ~&~$-$ ~&~$-$~& ~$+$ ~ & ~$-$~&~$+$
\\\hline
~$U(1)_{B-3L_\alpha}$& ~$-3$~ & ~$-3$~ &~$0$ ~&~$0$  & ~$0$ ~ &~$0$ ~&~$0$ ~&~$0$ ~&~$-3$~& ~$0$ ~ & ~$0$ ~& ~$3$
\\\hline \hline
\end{tabular}
\caption{Particle content and their charge assignment under $SU(2)_L\times U(1)_Y\times U(1)_{B-3L_{\alpha}}\times Z_2$ with $(\alpha,\beta,\gamma)$ a permutations of $(e,\mu,\tau)$.}
\label{TB:Charge}
\end{table}

\end{center}
\begin{figure}
\begin{center}
\includegraphics[width=0.4\linewidth]{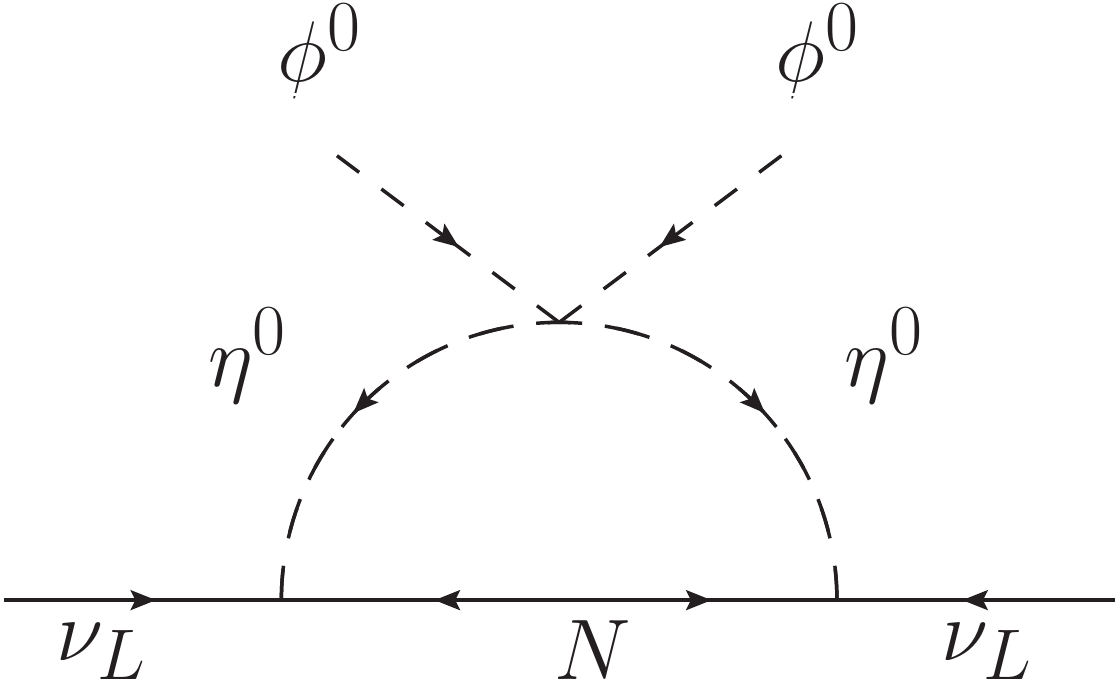}
\end{center}
\caption{Radiative neutrino mass at one loop.
\label{Fig:ma}}
\end{figure}
We extend the scotogenic model proposed by Ma\cite{Ma:2006km} to be compatible with the $U(1)_{B-3L_{\alpha}}$ gauge symmetry. The particle content related to lepton mass generation is listed in Table.I. In the model, three RH fermion singlets $N_{Ri}(i=1,2,3)$
and an inert scalar doublet field $\eta$ are added to the SM.  Among three $N_{R}$ fields, one of them carries the same non-zero charge as SM lepton doublet $L_\alpha$ under $U(1)_{B-3L_\alpha}$ gauge symmetry. In addition to the SM Higgs field $\Phi$, a singlet scalar $S$ is added to develop the vacuum expectation value (VEV) $v_S/\sqrt{2}$ after the spontaneous breaking of  $U(1)_{B-3L_{\alpha}}$ symmetry. The unbroken discrete $Z_{2}$ symmetry forbid the tree-level neutrino Yukawa interaction and assure the stability of DM.
In the original scotogenic model, the interactions relevant to lepton mass generation are given by
\begin{equation}
-\mathcal{L}~\supset~ f_{\alpha\beta}\bar{L}_{\alpha}\Phi \ell_{\beta R}+h_{\alpha i}\bar{L}_{\alpha}\eta^c N_{Ri}+\frac{M_{Nij}}{2} N_{Ri} N_{Rj}+\frac{1}{2}\lambda (\Phi^\dagger \eta )^2+\textsl{h.c}.
\end{equation}
The mass matrix $M_{N}$ can be diagonalized by unitary matrix $V$ satisfying
\begin{equation}
V^{T}M_{N}V=\hat{M}_{N}\equiv\text{diag}(M_{N_1},M_{N_2},M_{N_3}).
\end{equation}
The neutrino mass is generated radiatively, as show in Fig.~\ref{Fig:ma},  and can be computed exactly, i.e.
\begin{equation}
(M_{\nu})_{\alpha\beta}=\frac{1}{32\pi^{2}}\sum_{k}
h_{\alpha i}V_{ik}h_{\beta j}V_{jk}M_{Nk}\Big[\frac{m_{R}^{2}}{m_{R}^{2}\!-M_{Nk}^{2}}
\log\big(\frac{m_{R}^{2}}{M_{Nk}^{2}}\big)-\frac{m_{I}^{2}}{m_{I}^{2}\!-M_{Nk}^{2}}
\log\big(\frac{m_{I}^{2}}{M_{Nk}^{2}}\big)\Big]\\
\label{neutrinomass}\end{equation}
where $m_{0}^2\equiv(m_{R}^2+m_{I}^2)/2$.  In the so-called radiative seesaw scenario, the usual assumption of $m_{R}^2, m_{I}^2\ll M_{Nk}^2$ is adopted. Then the neutrino mass matrix in Eq.\eqref{neutrinomass} simplifies as
\begin{equation}
(M_{\nu})_{\alpha\beta}\simeq-\frac{\lambda v^2}{32\pi^{2}}\sum_{k}\frac{h_{\alpha i}V_{ik}h_{\beta j}V_{jk}}{M_{k}}\Big[\log\big(\frac{M_k^2}{m_0^2}\big)-1\Big]
\end{equation}
where $m_{R}$ and $m_{I}$ are the masses of $\Re \eta^0/\sqrt{2}$ and  $\Im \eta^0/\sqrt{2}$. In this paper, we focus on another interesting scenario given in Ref.\cite{Ma:2006km}. i.e $m_{R}^2, m_{I}^2\gg M_{Nk}^2$. In this case, the expression of neutrino mass matrix is given by
\begin{equation}\label{scale}\begin{split}
(M_{\nu})_{\alpha\beta}&\simeq -\frac{1}{32\pi^{2}}\frac{\lambda v^2}{m_{0}^2}\sum_{k}h_{\alpha i}V_{ik}h_{\beta j}V_{jk}M_{Nk}\\
&=-\frac{1}{32\pi^{2}}\frac{\lambda v^2}{m_{0}^2}(hM_{N}h^{T})_{\alpha\beta}
\end{split}
\end{equation}
Now we demonstrate how the one-zero-texture structures of $M_\nu$ arise from the
charge assignments given by Table. I. Taking $U(1)_{B-3L_{\tau}}$ gauge symmetry as example, the Yukawa matrix $h$, charged lepton mass matrix $M_l$ and the mass matrix of $N_R$ are given by
\begin{equation}
h= \left(\begin{array}{ccc}
h_{e1}&h_{e2}&0\\
h_{\mu1}&h_{\mu2}&0\\
0&0&h_{\tau3}
  \end{array}\right) \quad\quad
M_l= \frac{v}{\sqrt{2}}\left(\begin{array}{ccc}
f_{ee}&f_{e\mu}&0\\
f_{\mu e}&f_{\mu\mu}&0\\
0&0&f_{\tau\tau}
 \end{array}\right) \quad\quad
 M_{N}=\left(\begin{array}{ccc}
M_{11}& M_{12}&  M_{13}\\
M_{12} &M_{22}& M_{23}\\
M_{13}&M_{23}&0
  \end{array}\right)\label{yukawa}\end{equation}
Note that the $(1,3)$ and $(2,3)$ entries in $M_N$ originate from the interactions of $y_{13}N_{R1}N_{R3}S$ and $y_{23}N_{R2}N_{R3}S$.
Given the results in Eq.\eqref{yukawa} and using Eq.\eqref{scale}, $M_\nu$ can be written in the form
\begin{equation}\label{mlmv}
M_{\nu}= \left(\begin{array}{ccc}
(M_{\nu})_{11}&(M_{\nu})_{12}&(M_{\nu})_{13}\\
(M_{\nu})_{12}&(M_{\nu})_{22} &(M_{\nu})_{23}\\
(M_{\nu})_{13}&(M_{\nu})_{23}&0
  \end{array}\right).
\end{equation}
where
\begin{eqnarray}
&(M_{\nu} )_{11}=h_{e1}^2M_{11}+2h_{e1}h_{\mu 1}M_{12}+h_{\mu 1}^2M_{22}\\ \nonumber
&(M_{\nu} )_{12}=h_{e1}h_{e2}M_{11}+(h_{e2}h_{\mu 1}+h_{e1}h_{\mu 2})M_{12}+h_{\mu 1}h_{\mu 2}M_{22}\\ \nonumber
&(M_{\nu} )_{13}=h_{\tau3}(h_{e1}+h_{\mu 1})M_{13}\\ \nonumber
&(M_{\nu} )_{22}=h_{e2}^2M_{11}+2h_{e2}h_{\mu 2}M_{12}+h_{\mu 2}^2M_{22}\\ \nonumber
&(M_{\nu} )_{23}=h_{\tau3}(h_{e2}+h_{\mu 2})M_{23}
\end{eqnarray}
in the unit of $-\lambda v^2/32\pi^2m_0^2$.
The result given in Eq.\eqref{mlmv} directly reflects the one-zero-texture structure of the Majorana neutrino mass matrix. 
To have a more clear relation, the tedious expressions in Eq.\eqref{mlmv} would be further simplified with diagonal Yukawa matrix $h=\text{diag}(h_{e1},h_{\mu2},h_{\tau3})$, then
\begin{equation}\label{eq:TZs}
M_{\nu}= \left(\begin{array}{ccc}
 h_{e1}^2M_{11}& h_{e1} h_{\mu 2} M_{12} & h_{e1} h_{\tau 3} M_{13} \\
 h_{e1} h_{\mu 2} M_{12} & h_{\mu 2}^2M_{22} & h_{\mu2} h_{\tau 3} M_{23} \\
 h_{e1} h_{\tau 3} M_{13} & h_{\mu2} h_{\tau 3} M_{23} & 0
  \end{array}\right),
\end{equation}
in the unit of $-\lambda v^2/32\pi^2m_0^2$. The diagonal Yukawa matrix $h$ can be obtained by imposing a discrete $Z_3$ symmetry. The corresponding $Z_3$ charges are $L_e,e_R,N_{R1},S\sim \omega$, $L_\mu,\mu_R,N_{R2}\sim \omega^2$, and $L_\tau,\tau_R,N_{R3}\sim 0$ with $\omega=\text{exp}(i2\pi/3)$. Therefore, the form of $M_\nu$ is proportional to $M_N$ with an equal Yukawa structure $h_{e1}=h_{\mu2}=h_{\tau3}$.
 
The other two textures, predicted by $U(1)_{B-3L_e}$ and $U(1)_{B-3L_\mu}$ gauge symmetries, are produced in a similar approach. We list all of them in Table II.
The $M_{l}$ and $M_\nu$ are diagonalized by unitary matrix $V_{l}$ and $V_{\nu}$
\begin{equation}
M_{l}=V_{l}M_{l}^DV_{l}^{\dagger}\quad\quad M_{\nu}=V_{\nu}M_{\nu}^DV_{\nu}^{T}
\end{equation}
where $M_{l}^{D}\equiv\text{diag}(m_e,m_\mu,m_\tau)$ and $M_{\nu}^{D}\equiv\text{diag}(m_1,m_2,m_3)$. Then the Pontecorvo-Maki-Nakagawa-Sakata matrix $U_{\text{PMNS}}$\cite{PMNS} is given by
\begin{equation}
U_{PMNS}=V_{l}^{\dagger} V_{\nu}
\end{equation}
and can be parametrized as
\begin{align}
U_{\text{PMNS}}\! =\! \left(
\begin{array}{ccc}
c_{12} c_{13} & s_{12} c_{13} & s_{13}e^{-i\delta} \\
-s_{12}c_{23}-c_{12}s_{23}s_{13}e^{i\delta} & c_{12}c_{23} -s_{12}s_{23}s_{13}e^{i\delta} & s_{23}c_{13}\\
s_{12}s_{23}-c_{12}c_{23}s_{13}e^{i\delta} & -c_{12}s_{23}-s_{12}c_{23}s_{13}e^{i\delta} & c_{23}c_{13}
\end{array}
\right)\!\times\!
\text{diag}(e^{2i \rho},e^{2i\sigma},1)
\end{align}
Then we have
\begin{equation}
V_{\nu}=V_lU_{PMNS}
\label{vnu}\end{equation}
In terms of the zero-texture condition $(M_{\nu})_{\alpha\beta}=0$, we obtain the constraint condition
equation
\begin{equation}
\sum_{i=1,2,3}(V_{\nu})_{\alpha i}(V_{\nu})_{\beta i}m_{i}=0.
\label{sum}\end{equation}
From Table. II, one can see that the charged lepton matrix $M_l$ is nondiagonal. However, the degrees of freedom in $V_l$ do not change neutrino parameter space determined by one-zero texture in $M_\nu$. To see this, let's take $B-3L_\tau$ case as an example, which, from Eq.\eqref{sum}, satisfies the condition:
\begin{equation}
(V_{\nu})_{\tau 1}(V_{\nu})_{\tau 1}m_{1}+(V_{\nu})_{\tau 2}(V_{\nu})_{\tau 2}m_{2}+(V_{\nu})_{\tau 3}(V_{\nu})_{\tau 3}m_{3}=0
\label{sum2}\end{equation}
Because of the $B-3L_\tau$ symmetry, the charged lepton of $\tau$ do not mix with the ones of $e$ and $\mu$, leading to
 \begin{equation}
(V_\nu)_{\tau i}=(U_{PMNS})_{\tau i}
\label{veu}\end{equation}
By simply replacing the $(V_\nu)_{\tau i}$ in Eq. \eqref{sum2} with $(U_{PMNS})_{\tau i}$, the neutrino parameter space is constrained uniquely by the one-zero texture of $M_\nu$.
\begin{center}
\begin{table}
\begin{tabular}{|c|c|c|c|}\hline\hline
Pattern &Group & Texture of $M_{l}$ &Texture of $M_{\nu}$  \\
\hline
A& ~$U(1)_{B-3L_{e}}$
& ~ $\left(\begin{array}{ccc}
\times&0&0\\
0&\times&\times\\
 0&\times&\times
  \end{array}\right)$&
$ \left(\begin{array}{ccc}
0&\times&\times\\
\times&\times&\times\\
 \times&\times&\times
  \end{array}\right) $
\\ \hline
B& ~$U(1)_{B-3L_{\mu}}$
  &~$ \left(\begin{array}{ccc}
\times&0&\times\\
0&\times&0\\
 \times&0&\times
  \end{array}\right) $ &
$\left(\begin{array}{ccc}
\times&\times&\times\\
\times&0&\times\\
 \times&\times&\times
  \end{array}\right) $
  \\\hline
C& ~$U(1)_{B-3L_{\tau}}$
  &~ $\left(\begin{array}{ccc}
\times&\times&0\\
\times&\times&0\\
0&0&\times
  \end{array}\right)$  &
$ \left(\begin{array}{ccc}
\times&\times&\times\\
\times&\times&\times\\
\times&\times&0
  \end{array}\right) $
\\\hline \hline
\end{tabular}
\caption{Possible mass textures of charged leptons ($M_{l}$) and neutrinos ($M_{\nu}$), where $\times$ denotes a non-zero entry.}
\label{TB:Tex}
\end{table}
\end{center}
\section{Neutrino Mass and Mixings}
The systematic numerical analysis
of one-zero-texture structures was first presented in Ref. \cite{Lashin:2011dn}. In this section, we update
the predictions with latest global-fit results\cite{Esteban:2018azc} of neutrino oscillation parameters at $3\sigma$ confidence level (CL):

normal mass hierarchy (NH):
\begin{eqnarray}
\sin^{2}\theta_{12}/10^{-1}\in [2.75,3.50],\quad \sin^{2}\theta_{23}/10^{-1}\in [4.28,6.24],\quad \sin^{2}\theta_{13}/10^{-2}\in [2.044,2.437]\\ \nonumber
\delta m^2\equiv m_2^2-m_1^2\in [6.79, 8.01] \times 10^{-5} eV^2,\quad \Delta m^2\equiv |m_3^2-m_1^2|\in [2.431, 2.622] \times 10^{-3} eV^2
\end{eqnarray}

inverted mass hierarchy (IH):
\begin{eqnarray}
\sin^{2}\theta_{12}/10^{-1}\in [2.75,3.50],\quad \sin^{2}\theta_{23}/10^{-1}\in [4.33,6.23],\quad \sin^{2}\theta_{13}/10^{-2}\in [2.607,2.461]\\ \nonumber
\delta m^2\equiv m_2^2-m_1^2\in [6.79, 8.01] \times 10^{-5} eV^2,\quad \Delta m^2\equiv m_3^2-m_2^2\in [2.413, 2.606] \times 10^{-3} eV^2
\end{eqnarray}
Solving Eq. \eqref{sum} and using Eq. \eqref{veu}, we obtain two ratios of neutrino mass eigenvalues
\begin{eqnarray}
\kappa_{12}\equiv\frac{m_{1}}{m_2}=\frac{Re(U_{a3}U_{b3})Im(U_{a2}U_{b2}e^{2i\sigma})
-Re(U_{a2}U_{b2}e^{2i\sigma})Im(U_{a3}U_{b3})}{Re(U_{a1}U_{b1}e^{2i\rho})Im(U_{a3}U_{b3})
-Re(U_{a3}U_{b3})Im(U_{a1}U_{b1}e^{2i\rho})}
\nonumber \\
\kappa_{23}\equiv\frac{m_{2}}{m_3}=\frac{Re(U_{a1}U_{b1}e^{2i\rho})Im(U_{a3}U_{b3})
-Re(U_{a3}U_{b3})Im(U_{a1}U_{b1}e^{2i\rho})}{Re(U_{a2}U_{b2}e^{2i\sigma})Im(U_{a1}U_{b1}e^{2i\rho})
-Re(U_{a1}U_{b1}e^{2i\rho})Im(U_{a2}U_{b2}e^{2i\sigma})}
\label{kappa}\end{eqnarray}
The results of Eq. \eqref{kappa} imply that the two mass ratios ($\kappa_{12}$ and $\kappa_{13}$) are fully determined in terms of six neutrino mix parameters ($\theta_{12}, \theta_{23}, \theta_{13}, \delta, \rho, \sigma$). We further define the ratio of squared mass difference:
\begin{equation}
R_{\nu}\equiv \frac{\delta m^2}{|\Delta m^2|}
\label{rv}\end{equation}
The three neutrino masses $m_{1,2,3}$ are obtained from Eq.\eqref{kappa}:
\begin{equation}
m_{2}=\sqrt{\frac{\delta m^2}{1-\kappa_{12}^2}}\quad\quad m_1=\kappa_{12}m_2\quad\quad m_{3}=\frac{m_2}{\kappa_{23}}
\label{mass}\end{equation}
Then the $R_{\nu}$ defined in Eq.\eqref{rv}  can be rewritten in terms of $\kappa_{12}$ and $\kappa_{23}$
\begin{equation}
R_{\nu}=\frac{\kappa_{23}^2(1-\kappa_{12}^2)}{1-\kappa_{12}^2\kappa_{23}^2}
\label{rvnh}\end{equation}
for normal mass hierarchy and
\begin{equation}
R_{\nu}=\frac{\kappa_{23}^2(1-\kappa_{12}^2)}{\kappa_{23}^2-1}
\label{rvih}\end{equation}
for inverted mass hierarchy.

We now perform a numerical analysis for each texture structure in Table. II.
A set of random number inputs is generated for the
three mixing angles $(\theta_{12}, \theta_{23}, \theta_{13})$ in their $3\sigma$ range. Instead, we generate a random input of ($\delta, \rho, \sigma$) in the range of $[0, 2\pi)$.
From Eqs.\eqref{rvnh} and \eqref{rvih}, $R_{\nu}$ is fully determined by the neutrino parameters of ($\theta_{12},\theta_{23},\theta_{13},\delta,\rho,\sigma)$.
We require the input
scattering points acceptable only when $R_{\nu}$ falls inside the
$3\sigma$ range $[\delta m^{2}_{min}/\Delta m^{2}_{max},\delta
m^{2}_{max}/\Delta m^{2}_{min}]$. From
Eq.\eqref{mass}, the three absolute
scale of neutrino masses $m_{1,2,3}$ are obtained. We should also consider the robust bound on the sum of three neutrino masses $\Sigma_\nu(\equiv \Sigma m_i)<0.12$ eV set by Planck Collaboration at $95\%$ CL\cite{Aghanim:2018eyx}. However a latest analysis based on the physically motivated neutrino mass models yields a looser upper bound $\Sigma_\nu<0.26$ eV. at $95\%$ CL\cite{Loureiro:2018pdz}. In the following analysis, we check the consistency with the both of the bounds. In addition to $\Sigma_\nu$, we also calculate the effective Majorana neutrino mass
\begin{equation}
m_{ee}=\left|m_{1}c_{12}^{2}c_{13}^{2}+m_{2}s_{12}^{2}c_{13}^{2}e^{2i\rho}+m_{3}s_{13}^{2}e^{2i\sigma}\right|
\label{mee}\end{equation}
which can be explored by
neutrinoless double beta decay ($0\nu\beta\beta$) experiments. Among the current experiments in operation, the KamLAND-Zen Collaboration\cite{KamLAND-Zen:2016pfg} has set the most stringent limit of $m_{ee}< (0.061-0.165)$ eV  depending on the nuclear matrix element value of the $^{136}$Xe decays, while next-generation experiments aim for an improved sensitivity of $ m_{ee}$ being up to 0.01 eV.

\begin{figure}
\begin{center}
\includegraphics[width=1.05\linewidth]{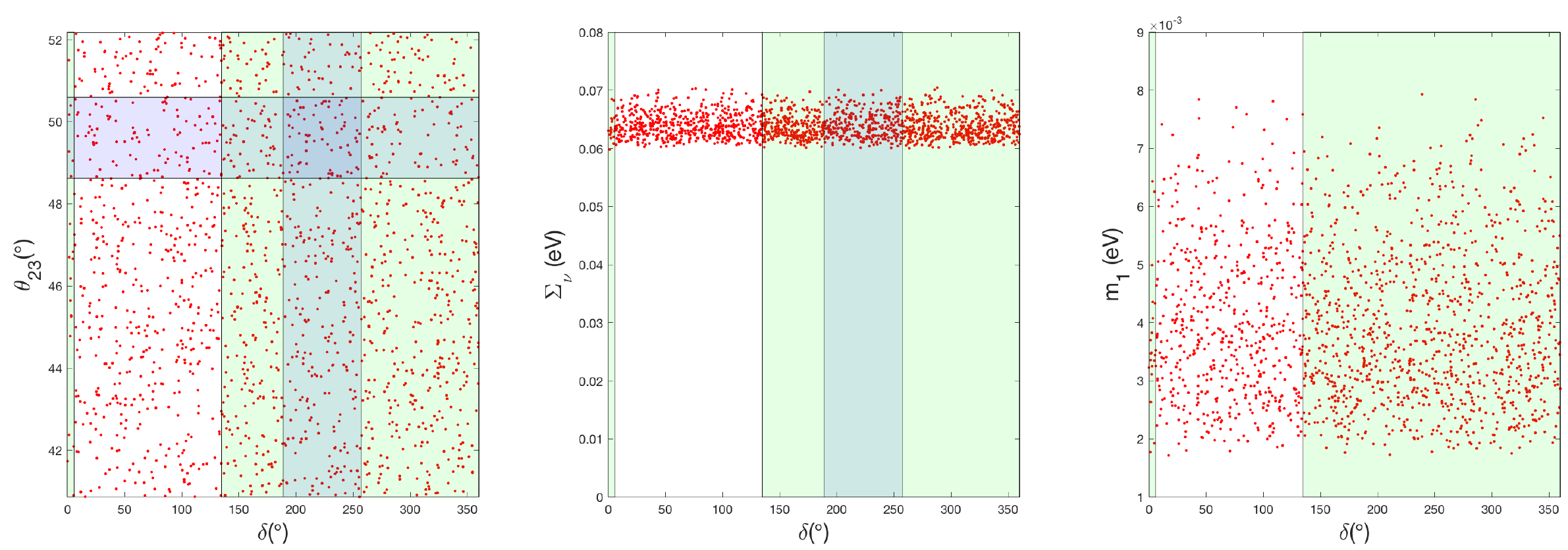}
\end{center}
\caption{Allowed samples of texture A in the case of normal mass hierarchy. The light green bands represent the $3\sigma$ uncertainty of $\delta$. The light blue bands represent the $1\sigma$ uncertainty of $\delta$ or $\theta_{23}$. \label{eeNH} }
\end{figure}

\begin{figure}
\begin{center}
\includegraphics[width=1.05\linewidth]{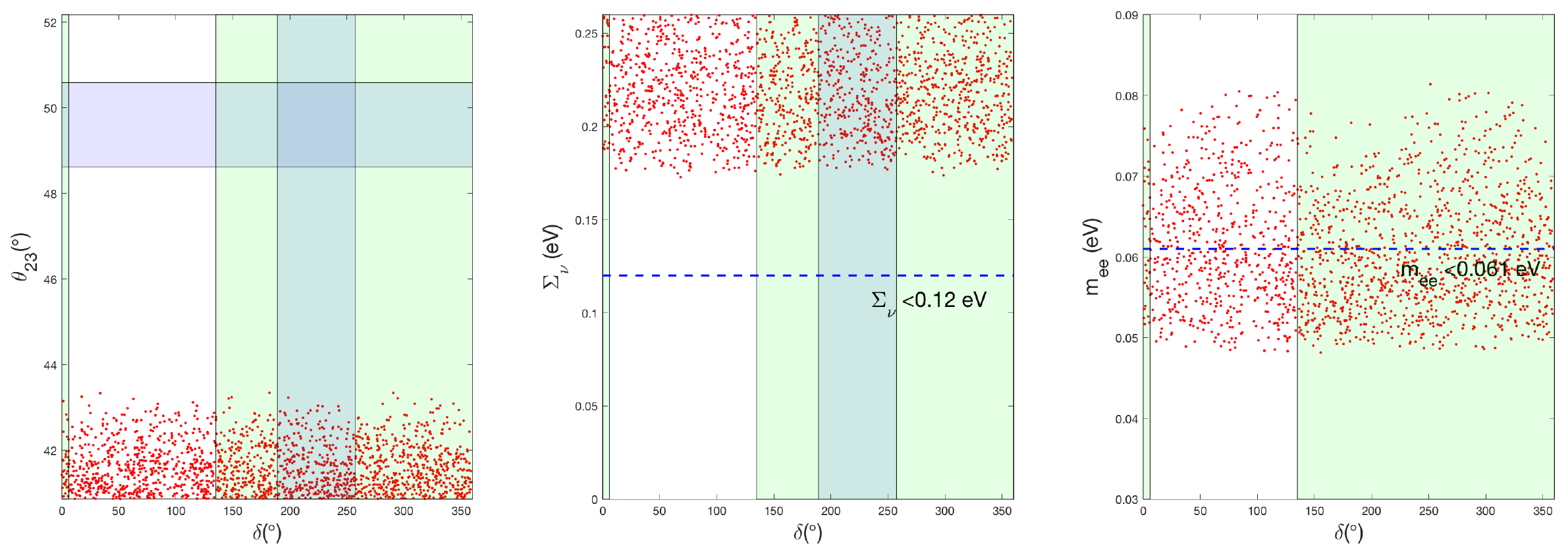}
\end{center}
\caption{Allowed samples of texture B in the case of normal mass hierarchy.
The horizontal blue dashed line in the middle panel is the upper bound imposed by the Planck Collaboration: $\sum_{\nu} m_{i}<0.12$ eV. The horizontal blue dashed line in the right panel is the upper limit of $m_{ee}$ set by the KamLAND-Zen Collaboration. \label{mmNH} }
\end{figure}

\begin{figure}
\begin{center}
\includegraphics[width=1.1\linewidth]{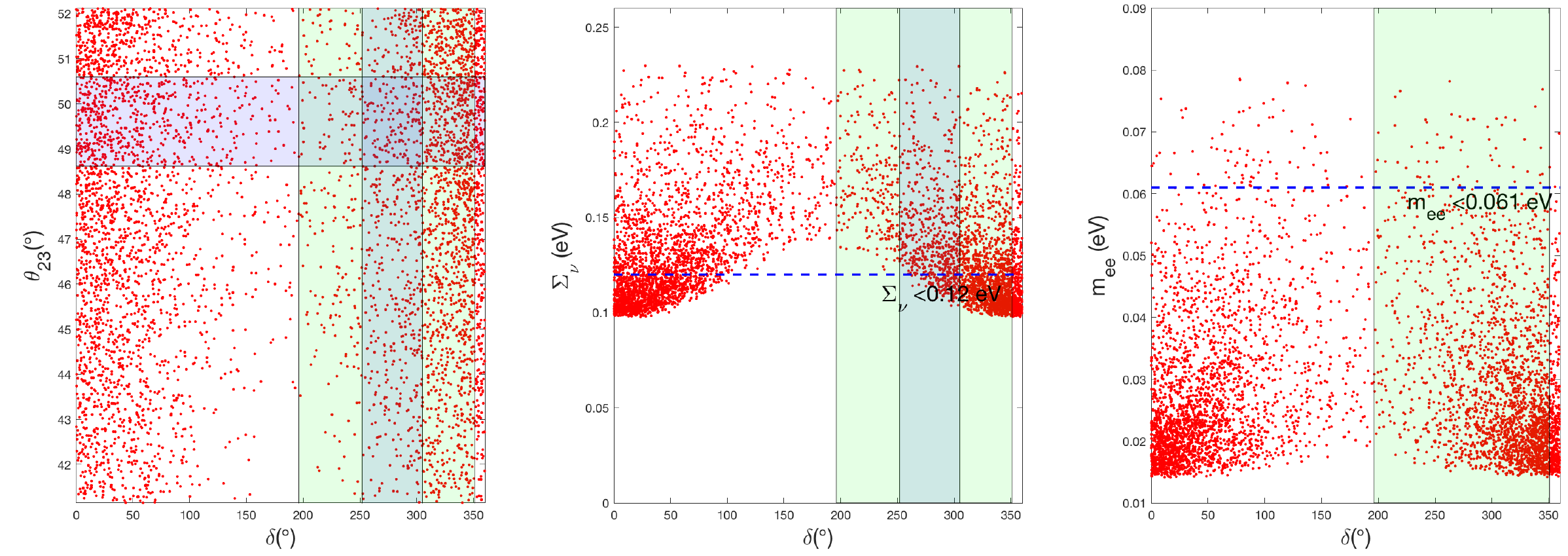}
\end{center}
\caption{Allowed samples of texture B in the case of inverted mass hierarchy.  \label{mmIH}}
\end{figure}

\begin{figure}
\begin{center}
\includegraphics[width=1.1\linewidth]{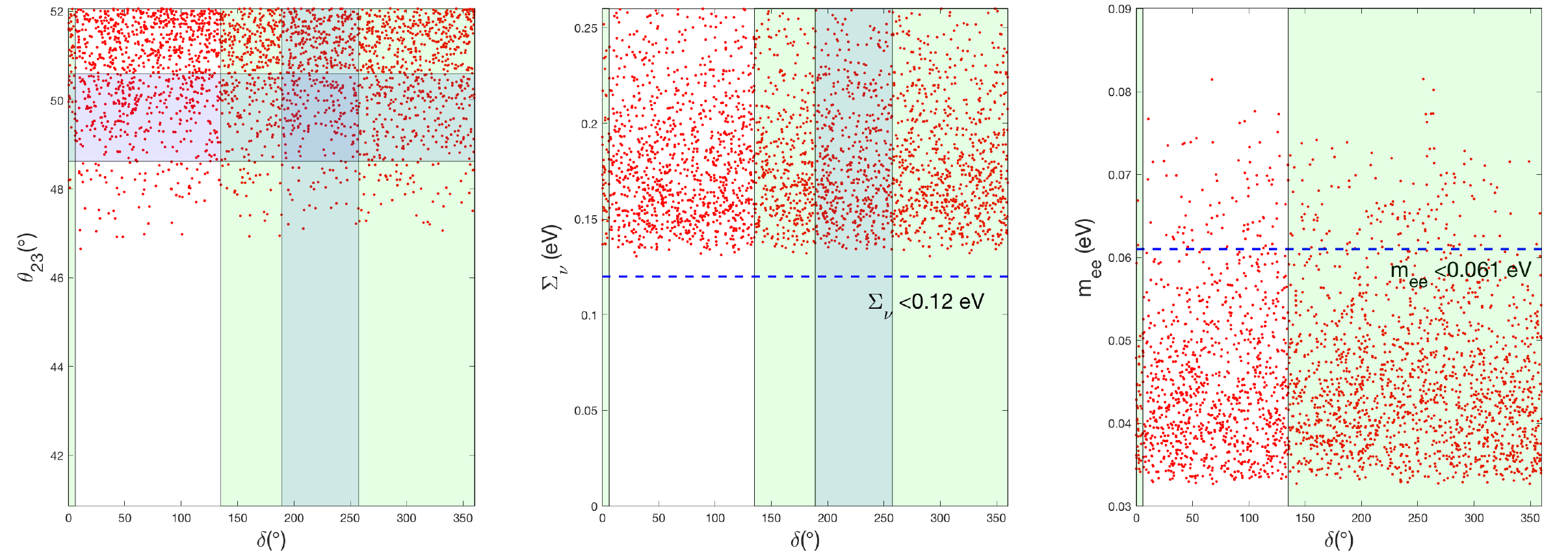}
\end{center}
\caption{Allowed samples of texture C in the case of normal mass hierarchy. \label{ttNH}}
\end{figure}

\begin{figure}
\begin{center}
\includegraphics[width=1.05\linewidth]{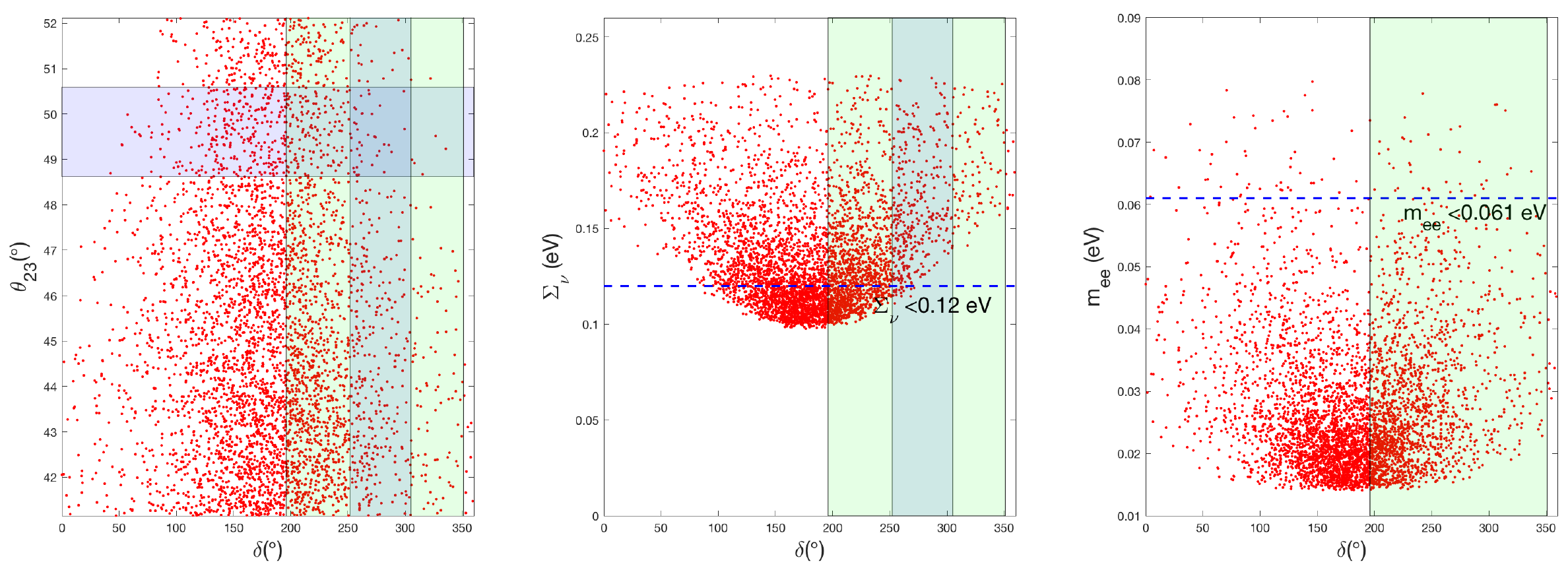}
\end{center}
\caption{Allowed samples of texture C in the case of inverted mass hierarchy. \label{ttIH}}
\end{figure}

We present the allowed range of neutrino oscillation parameters for each viable pattern in Figs.\ref{eeNH}-\ref{ttIH} The  $U(1)_{B-3L_e}$ gauge symmetry leads to pattern A of one-zero texture structure ( see Table. II). we find it phenomenologically allowed only for the case of normal mass hierarchy. Since $(M_\nu)_{ee} = 0$, the predicted effective Majorana neutrino mass $m_{ee}$ is exactly zero for the $0\nu\beta\beta$ decay. Figure.\ref{eeNH} shows no strong correlation between $(\delta, \theta_{23})$ and $\theta_{23}$ covers all $3\sigma$ allowed region. The mass spectrum admits a strong normal hierarchy with $\Sigma_\nu\simeq 0.06-0.07$ eV; meanwhile, a lower bound on the lowest neutrino mass $m_1\geq 0.0015$ eV is achieved.
The $B-3L_{\mu}$ gauge symmetry leads to pattern B. It is found that both normal mass hierarchy and inverted mass hierarchy are allowed to satisfy the neutrino oscillation
data, and corresponding scanning results are presented in Figs.\ref{mmNH} and \ref{mmIH}, respectively. In the case of normal mass hierarchy,  the predicted value of $\theta_{23}$ is restricted to be less than $\pi/4$ which, though allowed at $3\sigma$ level, is disfavored at $1\sigma$ level. Besides, the value of $\Sigma_{\nu}$ turns out to be relative large, which is excluded by the Planck 2018 limit.
In the case of inverted mass hierarchy, one can observe from the left panel of Fig. \ref{mmIH} that the sample points of ($\delta$, $\theta_{23}$) fully cover the $3\sigma$ and $1\sigma$ allowed region. Furthermore, the interesting correlation of $\Sigma_\nu$ versus $\delta$, shown in the middle panel of Fig. \ref{mmIH}, indicates that the Planck bound $\Sigma_\nu <0.12$ is satisfied only for $\delta\in[0^{\circ}, 110^{\circ}]\cup[250^{\circ},360^{\circ}]$, a range almost excluded at $1\sigma$ level. Both normal mass hierarchy and inverted mass hierarchy of pattern B predict the $m_{ee}=\mathcal{O}(0.01)$ eV, which is promising to be detected in the forthcoming experiments.
The $B-3L_{\tau}$ gauge symmetry leads to pattern C. In Fig.\ref{ttNH} and \ref{ttIH}, we present the allowed region of pattern C for normal mass hierarchy and inverted mass hierarchy respectively. Before proceeding, one notes that the $\mu-\tau$ permutation transformation can relate pattern B to pattern C by swapping the indices $e$ and $\tau$ of the entries and the neutrino oscillation parameters between $M_{\nu}$ (pattern B) and $\tilde{M}_\nu$ (pattern C) are given by
\begin{equation}
\theta_{12}=\tilde{\theta}_{12}\quad\quad \theta_{13}=\tilde{\theta}_{13}\quad\quad \theta_{23}=\frac{\pi}{2}-\tilde{\theta}_{23}\quad\quad \delta=\tilde{\delta}\pm \pi
\end{equation}
In the case of normal mass hierarchy, due to the $\mu-\tau$ symmetry, we arrive at the $\theta_{23}$ mixing angle greater than $\pi/4$, which is located in the $1\sigma$ range. Meanwhile, we obtain a rather large value of $\Sigma_\nu$ (Fig. \ref{ttNH}), as similar as pattern B, being disfavored by Planck 2018 results. We now discuss the pattern C with inverted mass hierarchy, which is particularly interesting for us. The left panel of Fig. \ref{ttIH} shows no obvious correlation between $\delta$ and $\theta_{23}$, while the correlation $(\delta, \Sigma_{\nu})$ shows a specific geometrical shape in the middle panel. In particular, part of the allowed region to satisfy the Planck 2018 bound  $\Sigma_\nu<0.12$ eV fully covers the $1\sigma$ range of Dirac-CP phase: $\delta \in [1.40\pi,1.68\pi]$. Therefore, for the inverted mass hierarchy of pattern C, there exists allowed regions of neutrino parameters in which all the experimental constraints are satisfied. Finally, both normal mass hierarchy and inverted mass hierarchy of pattern C predict the $m_{ee}$ with a lower bound greater than $0.01$ eV, which will be explored by the next generation experiments.

\section{Other Phenomenology}\label{Sec:PH}
In this section, we take the $U(1)_{B-3L_\mu}$ extension of scotogenic model for some other specific phenomenology studies. The results for $U(1)_{B-3L_{e,\tau}}$ are similar except for the corresponding gauge boson $Z'$ \cite{Han:2019zkz}. The $U(1)_{B-3L_\mu}$ is spontaneously broken by the $S$ VEV $\langle S \rangle=v_S/\sqrt{2}$ at TeV scale, and generates the mass for the new gauge boson $Z'$,
\begin{equation}
M_{Z'}=3g' v_S,
\end{equation}
where $g'$ is the gauge coupling of $U(1)_{B-3L_\mu}$. The LEP bound requires $M_{Z'}/g'\gtrsim7$ TeV \cite{Cacciapaglia:2006pk}. In unitary gauge, the neutral component of $Z_2$ even scalars are parameterized as
\begin{equation}
\Phi^0=\frac{v+\phi}{\sqrt{2}},~S=\frac{v_S+S^0}{\sqrt{2}}.
\end{equation}
The neutral scalars $\phi$ and $S^0$ are correlated with the mass eigenstates $h$ and $H$ via a rotation
\begin{align}
\left(
\begin{array}{c}
h\\
H
\end{array}\right)=\left(
\begin{array}{cc}
\cos\alpha & -\sin\alpha\\
\sin\alpha & \cos\alpha
\end{array}\right)\left(
\begin{array}{c}
\phi\\
S^0
\end{array}\right),
\end{align}
where the mixing angle $\alpha$ can be treated as a free parameter. To satisfy various theoretical and experimental constraints, $\sin\alpha\lesssim0.2$ should be satisfied \cite{Robens:2016xkb}.

\subsection{Lepton Flavor Violation}

It is well known that the Yukawa interaction $\bar{L}\eta^cN_R$ could induce lepton flavor violation process \cite{Kubo:2006yx}. Systematic studies on all LFV processes in scotogenic models have already been performed in Ref.~\cite{Toma:2013zsa}. Currently, MEG experiment has set the most stringent constraint on $\mu\to e\gamma$, which requires BR$(\mu\to e\gamma)<4.2\times10^{-13}$ at 90\% CL \cite{TheMEG:2016wtm}. The future upgrade of MEG could push the limit down to about $6\times10^{-14}$ \cite{Baldini:2013ke}. Meanwhile, the limits on LFV $\tau$ decays are much loose, i.e., BR$(\tau\to \mu\gamma)<4.4\times10^{-8}$ and BR$(\tau\to e\gamma)<3.3\times10^{-8}$ \cite{Aubert:2009ag}. Hence, large Yukawa coupling in the $\tau$ sector is viable if one imposes hierarchal Yukawa structure as $|h_{ei}|\lesssim|h_{\mu i}|\lesssim|h_{\tau i}|$ \cite{Vicente:2014wga}. On the other hand, if one consider universal Yukawa structure, then it is clearly expected that the tightest constraint comes from $\mu\to e\gamma$. The corresponding branching ratio is calculated as \cite{Toma:2013zsa}
\begin{equation}
\text{BR}(\ell_\alpha\to \ell_\beta\gamma) = \frac{3(4\pi^3)\alpha_\text{em}}{4G_F^2} |A_D|^2 \text{BR}(\ell_\alpha\to \ell_\beta \nu_\alpha \bar{\nu}_\beta),
\end{equation}
with the dipole form factor
\begin{equation}
A_D=\sum_{i=1}^3 \frac{h'_{\beta i} h'^*_{\alpha i}}{2(4\pi)^2 M^2_{\eta^+}}
F_2\left(\frac{M_{N_i}^2}{M^2_{\eta^+}}\right),
\end{equation}
where $h^\prime=hV_N$ is the Yukawa coupling in the mass eigenstates, and the loop function $F_2(x)$ is given by
\begin{equation}
F_2(x)=\frac{1-6x+3x^2+2x^3-6x^2\log x}{6(1-x)^4}.
\end{equation}

\begin{figure}
\begin{center}
\includegraphics[width=0.45\linewidth]{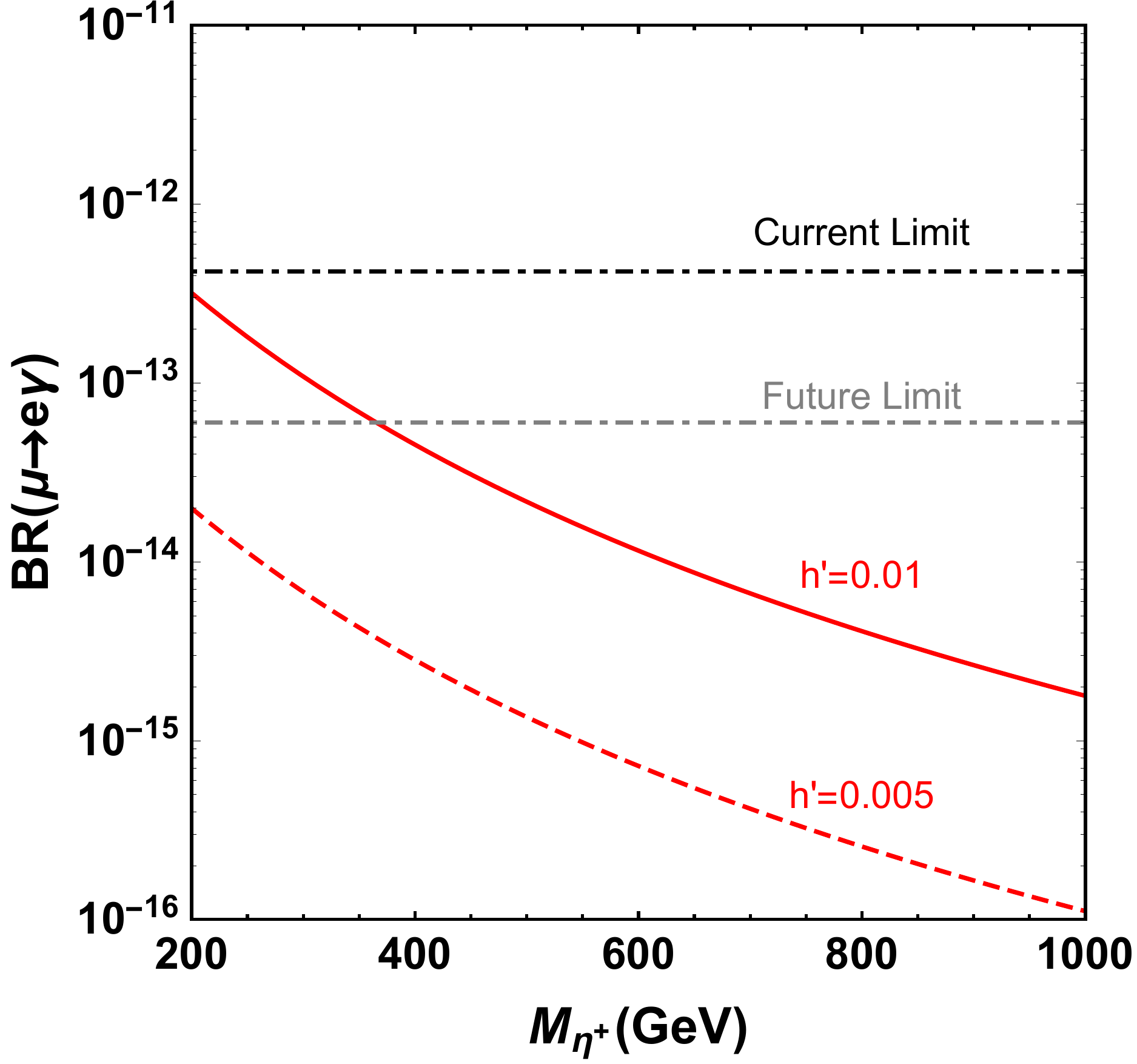}
\includegraphics[width=0.45\linewidth]{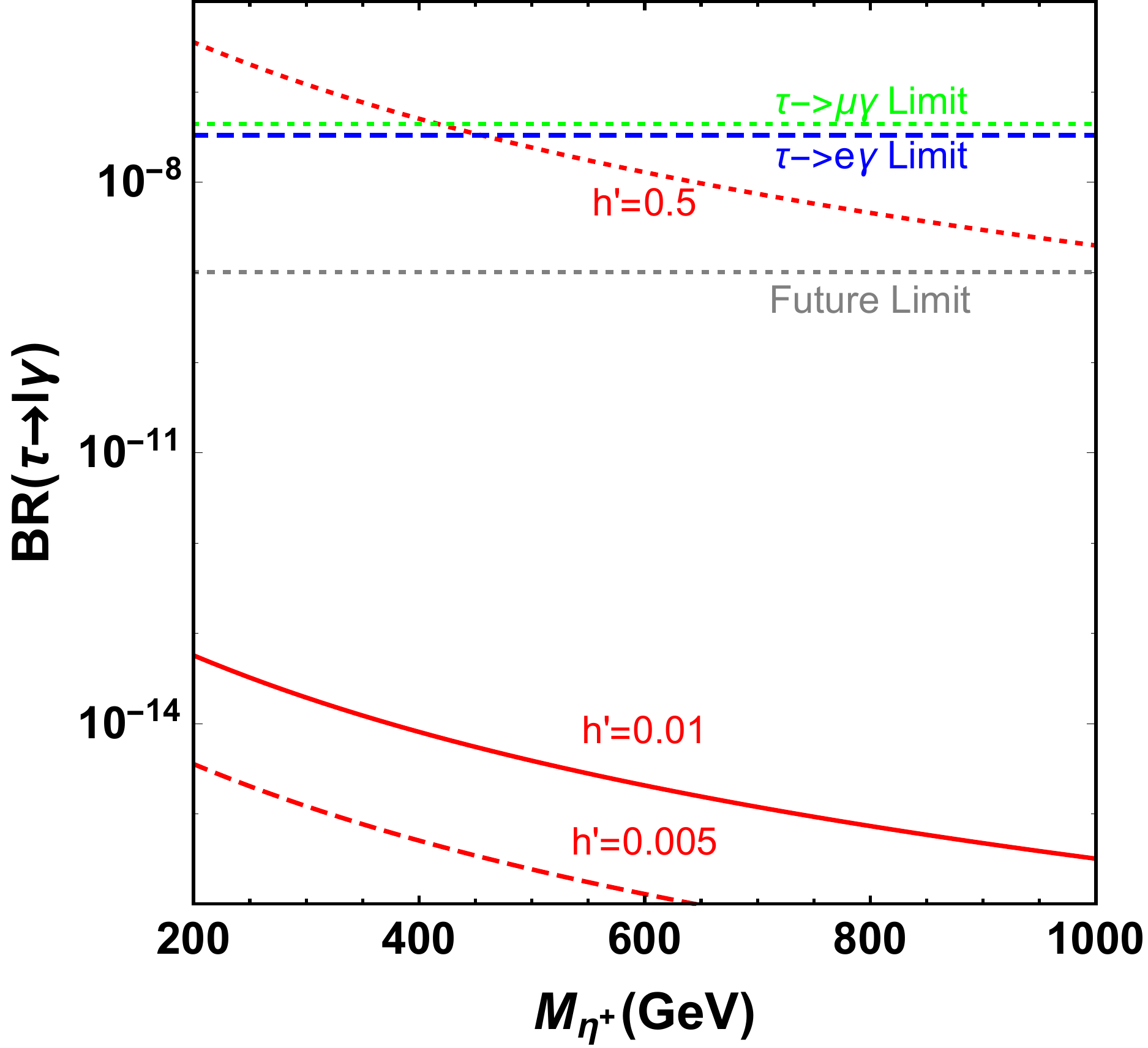}
\includegraphics[width=0.45\linewidth]{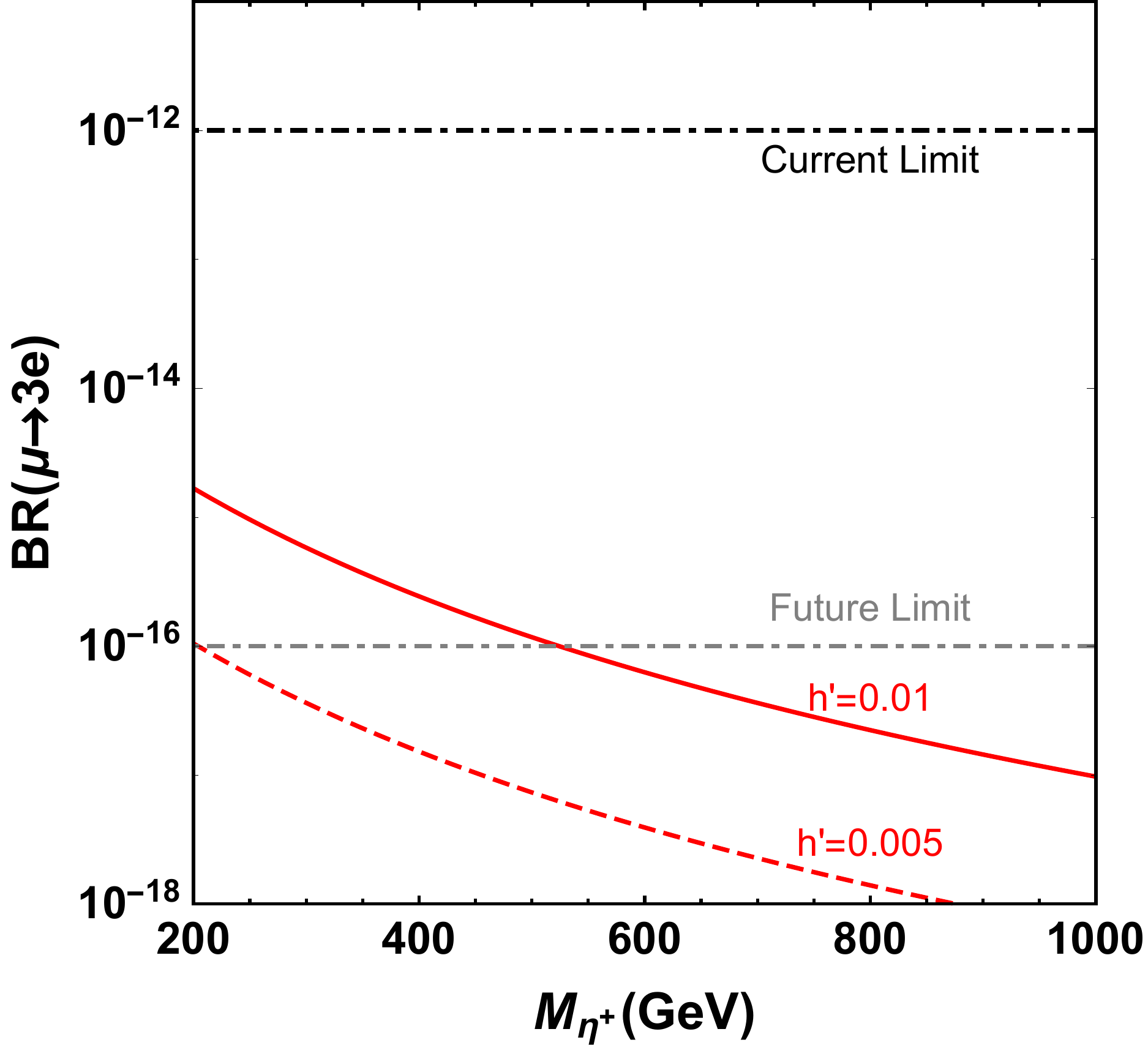}
\includegraphics[width=0.45\linewidth]{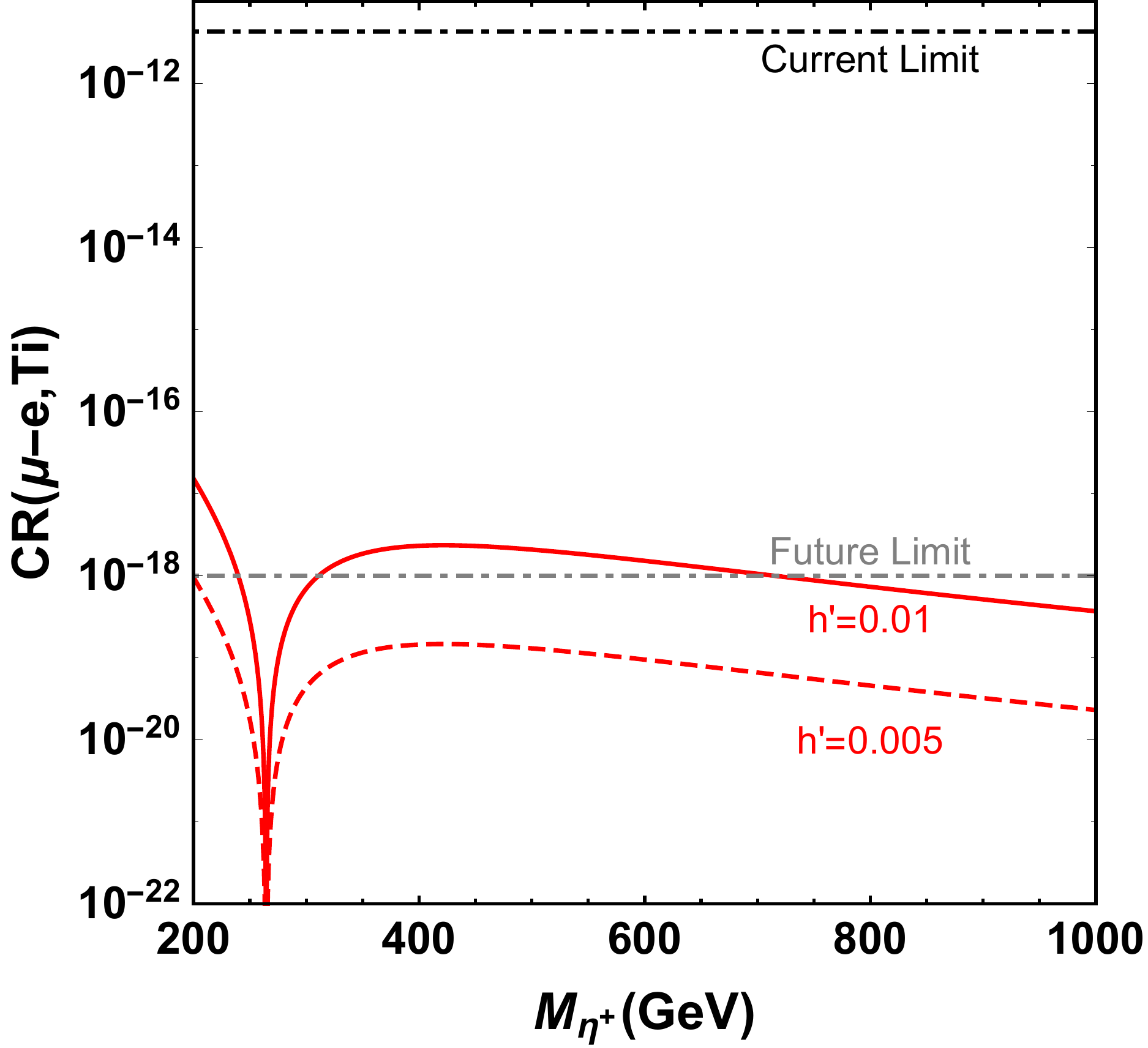}
\end{center}
\caption{Predicted results for various LFV processes as a function of $M_{\eta^+}$. Here, we assume $h'_{\alpha i}=h'$ and $M_{N_1}\simeq M_{N_{2,3}}=200$ GeV for illustration.
 \label{Fig:LFV}}
\end{figure}

The three-body decay $\ell_\alpha\to \ell_\beta \bar{\ell}_\beta\ell_\beta$ is another kind of LFV process. The current limit on BR($\mu\to3e$) is  $1.0\times10^{-12}$ \cite{Bellgardt:1987du} and is expected to reach $10^{-16}$ in the future \cite{Blondel:2013ia}. The limits on BR($\tau\to 3e$) and BR($\tau\to 3\mu$) are $2.7\times10^{-8}$ and $2.1\times10^{-8}$ \cite{Hayasaka:2010np}.
Considering the simplified scenario with small Yukawa coupling, e.g. $h'\lesssim0.01$, then the three-body decay $\ell_\alpha\to \ell_\beta \bar{\ell}_\beta\ell_\beta$ is dominant by the $\gamma$-penguins. The branching ratio is given by
\begin{eqnarray}
\text{BR}(\ell_\alpha\to \ell_\beta \bar{\ell}_\beta \ell_\beta) &=&
\frac{3 (4\pi)^2 \alpha_\text{em}^2}{8G_F^2}\bigg[ |A_{ND}|^2 +|A_D|^2\left(\frac{16}{3}\log\left(\frac{m_\alpha}{m_\beta}\right)
-\frac{22}{3}\right) \\ \nonumber
&&\hspace{5 em} - \big(2 A_{ND} A_D^*+\text{h.c.}\big)\bigg]\times \text{BR}(\ell_\alpha\to \ell_\beta \nu_\alpha \bar{\nu}_\beta),
\end{eqnarray}
with the non-dipole form factor
\begin{equation}
A_{ND}=\sum_{i=1}^3 \frac{h'_{\beta i} h'^*_{\alpha i}}{6(4\pi)^2 M^2_{\eta^+}}
G_2\left(\frac{M_{N_i}^2}{M^2_{\eta^+}}\right),
\end{equation}
and the loop function $G_2(x)$
\begin{equation}
G_2(x)=\frac{2-9x+18x^2-11x^3+6x^3\log x}{6(1-x)^4}.
\end{equation}
Note when the photonic dipole contribution dominates, a simple relation is derived \cite{Arganda:2005ji}
\begin{equation}
\text{BR}(\ell_\alpha\to \ell_\beta \bar{\ell}_\beta \ell_\beta)\simeq
\frac{\alpha_\text{em}}{3\pi}\left(\log\left(\frac{m_\alpha^2}{m_\beta^2}\right)
-\frac{11}{4}\right)\times \text{BR}(\ell_\alpha\to \ell_\beta\gamma).
\end{equation}
Therefore comparing with $\ell_\alpha \to \ell_\beta \gamma$,  $\ell_\alpha\to \ell_\beta \bar{\ell}_\beta \ell_\beta$ is suppressed. Due to quite loose limits on three body $\tau$ decays, we will focus on $\mu\to 3 e$ in following.

Meanwhile, the planed $\mu-e$ conversion in nuclei experiments will push current limits on the conversion rates, e.g., $\text{CR}(\mu-e,\text{Ti})<4.3\times10^{-12}$ \cite{Dohmen:1993mp} down to about $10^{-18}$ \cite{PRIME}. The converting rate is then expressed as \cite{Kitano:2002mt}
\begin{equation}
\text{CR}(\mu-e,\text{Ti})=2G_F^2\left|A_R D+\tilde{g}_{LV}^p V^p\right|^2,
\end{equation}
with
\begin{equation}
A_R=-\frac{\sqrt{2}}{8G_F}e A_D,\quad \tilde{g}_{LV}^p=\frac{\sqrt{2}}{G_F} e^2 A_{ND},
\end{equation}
and $D=0.0864 M_\mu^{5/2}$, $V^p=0.0396 M_\mu^{5/2}$. Due to relative opposite sign between above $A_R$ and $\tilde{g}_{LV}^p$ terms, the cancellation effect is expected.

The predicted results for LFV are shown in Fig.~\ref{Fig:LFV}. For electroweak scale $N_i$ and $\eta^+$, an universal Yukawa coupling $h'=0.01$ can escape current limit but is within future limit. Specifically speaking, the $\mu \to e\gamma$ process could probe $M_{\eta^+}\lesssim360$ GeV with $M_{N}=200$ GeV.  Although the $\mu\to 3e$ process is suppressed, the great improvement of future experiment will set a more tight bound than $\mu\to e\gamma$, i.e., excluding $M_{\eta^+}\lesssim 520$ GeV. The cancellation effect for $\mu-e$ conversion is obvious around $M_{\eta^+}\sim260$ GeV. Together with other LFV process, $\mu-e$ conversion could exclude $M_{\eta^+}\lesssim700$ GeV. As for the $\tau\to e\gamma/\mu\gamma$, the Yukawa coupling as large of $h'=0.5$ is still allowed, but is within future reach.
It is also clear that $h'=0.005$ is beyond all future experimental reaches.

\begin{figure}
\begin{center}
\includegraphics[width=0.45\linewidth]{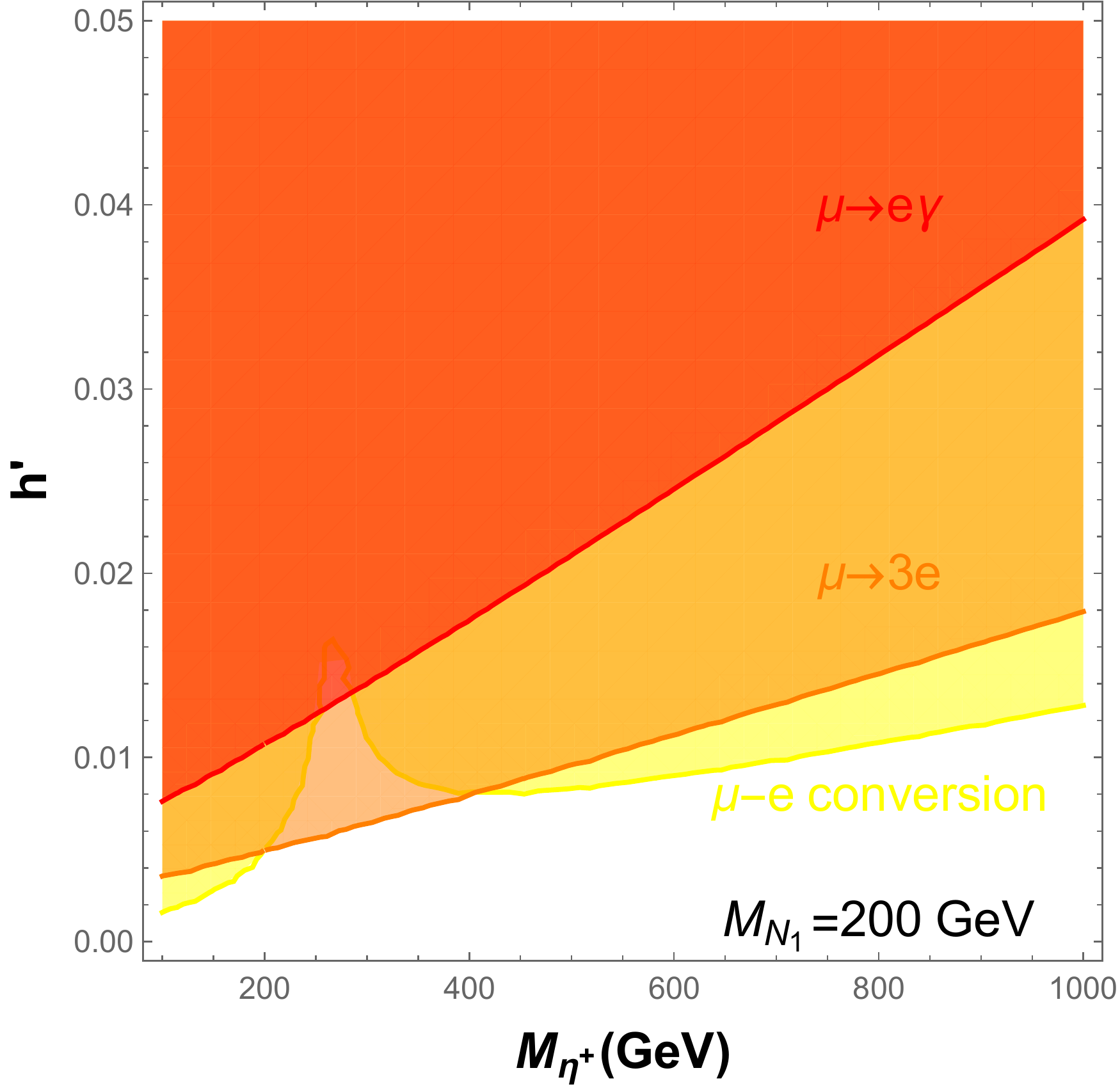}
\includegraphics[width=0.45\linewidth]{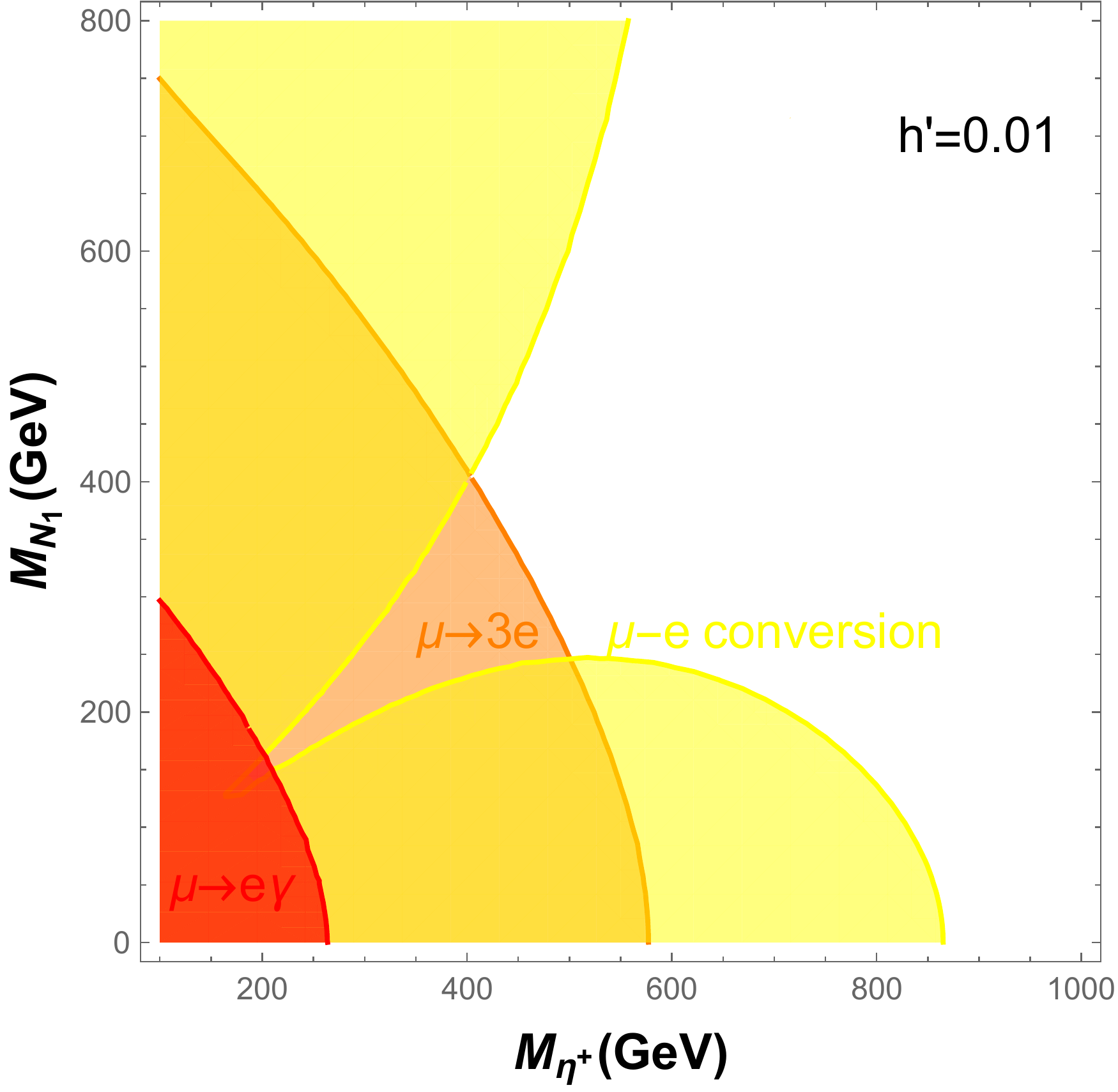}
\end{center}
\caption{Current and future prospect excluded region of various LFV processes in the $h'-M_{\eta^+}$ plane (left) and in the $M_{N_1}-M_{\eta^+}$ plane (right).
 \label{Fig:PS}}
\end{figure}

The derived current and future excluded region of various LFV processes are shown in Fig.~\ref{Fig:PS}, where we have assumed $h_{\alpha i}=h'$ for illustration.
The limits from $\mu\to e\gamma$ and $\mu\to 3e$ become less strict when $M_{\eta^+}$ is larger. For instance, current $\mu\to e\gamma$ sets the limit $h'\lesssim0.01$ when $M_{\eta^+}\sim200$ GeV and $h'\lesssim0.04$ when $M_{\eta^+}\sim1000$ GeV, and $\mu\to 3e$ will improve the limit by a factor of about two. As for $\mu-e$ conversion, large cancellation happens when $M_{N}\lesssim M_{\eta^+}$, and its limit is more stringent than $\mu\to 3 e$ when $M_{\eta^+}<200$ GeV or $M_{\eta^+}>400$ GeV. Then if we fix $h'=0.01$, $\mu\to e\gamma$ could exclude the region $M_{N_1}\lesssim300$ GeV and $M_{\eta^+}\lesssim250$ GeV. For EW-scale $N_1$, $\mu\to 3e$ ($\mu-e$ conversion) will probe $M_{\eta^+}\lesssim500 (700)$ GeV.

\subsection{Dark Matter}

The dark matter candidate in this model is $N_1$ under our consideration. As for annihilation of $N_1$ DM, there are three kinds of interactions
\begin{description}
  \item[$\eta$-portal] The same Yukawa interaction $ h'_{\alpha i} \bar{L}_\alpha\eta^c N_i$ inducing LFV is also involved in $N_1$ annihilation. Due to tight constraints on $h'$ from LFV, $N_1N_1$ pair annihilation is suppressed for small value of $h'$, which leads to an excess of observed relic density.  For instance, if we set $h'_{\alpha i}=0.01$, then the $\eta$-portal annihilation cross section
      \begin{equation}
      \langle\sigma v\rangle\simeq \sum_{\alpha\beta}
      \frac{|h'_{\alpha 1}h'^*_{\beta 1} |^2 r_1^2(1-2r_1+2r_1^2)}{24\pi M_{N_1}}\langle v^2\rangle\sim6\times10^{-33} \text{cm}^3 \text{s}^{-1},
      \end{equation}
      where $r_1=M_{N_1}^2/(M_\eta^2+M_{N_1}^2)$. Therefore, the contribution of $\eta$-portal annihilation is less than $10^{-7}$ in our consideration. One pathway to overcame this confliction is considering hierarchal Yukawa structure or co-annihilation mechanism \cite{Vicente:2014wga,Suematsu:2009ww}. Besides, it is also possible to employ the freeze-in production of $N_1$, which requires tiny Yukawa coupling $h'_{\alpha 1}\sim 10^{-10}$ \cite{Molinaro:2014lfa}.
  \item[$H$-portal] The Higgs portal interactions $N_1N_1H(h)$ can mediate $s$-channel $N_1N_1$ annihilation \cite{Okada:2010wd,Kanemura:2011vm}. Correct relic density can be easily obtained when $M_{N_1}\sim M_{H}/2$ for EW scale $M_H$ \cite{Ding:2018jdk}. And the LFV constraints are escaped by assuming small $h'\lesssim 0.01$. At the meantime, the Higgs portal interactions also induce spin-independent DM-nucleon scattering, which might be probed by DM direct detection experiments.
  \item[$Z'$-portal] Although only $N_{R3}$ is charged under the $U(1)_{B-3L_\mu}$, mixings of singlet fermions $N_R$ also induce an effective coupling of $N_1\gamma^\mu\gamma_5 N_1 Z'_\mu$. The $Z'$-portal interaction can also mediate $N_1N_1$ annihilation \cite{Okada:2016gsh,Okada:2016tci}. Meanwhile, direct LHC search for the dilepton signature $pp\to Z'\to \ell^+\ell^-$ requires that $Z'$ should be best above TeV scale \cite{Han:2017ars,Han:2018zcn}. Due to the Majorana nature of $N_1$, the DM-nucleon scattering cross section mediated by $Z'$ is suppressed by $v^2_\text{rel}\sim10^{-6}$ \cite{Baek:2015fea}.
\end{description}
\begin{figure}
\begin{center}
\includegraphics[width=0.45\linewidth]{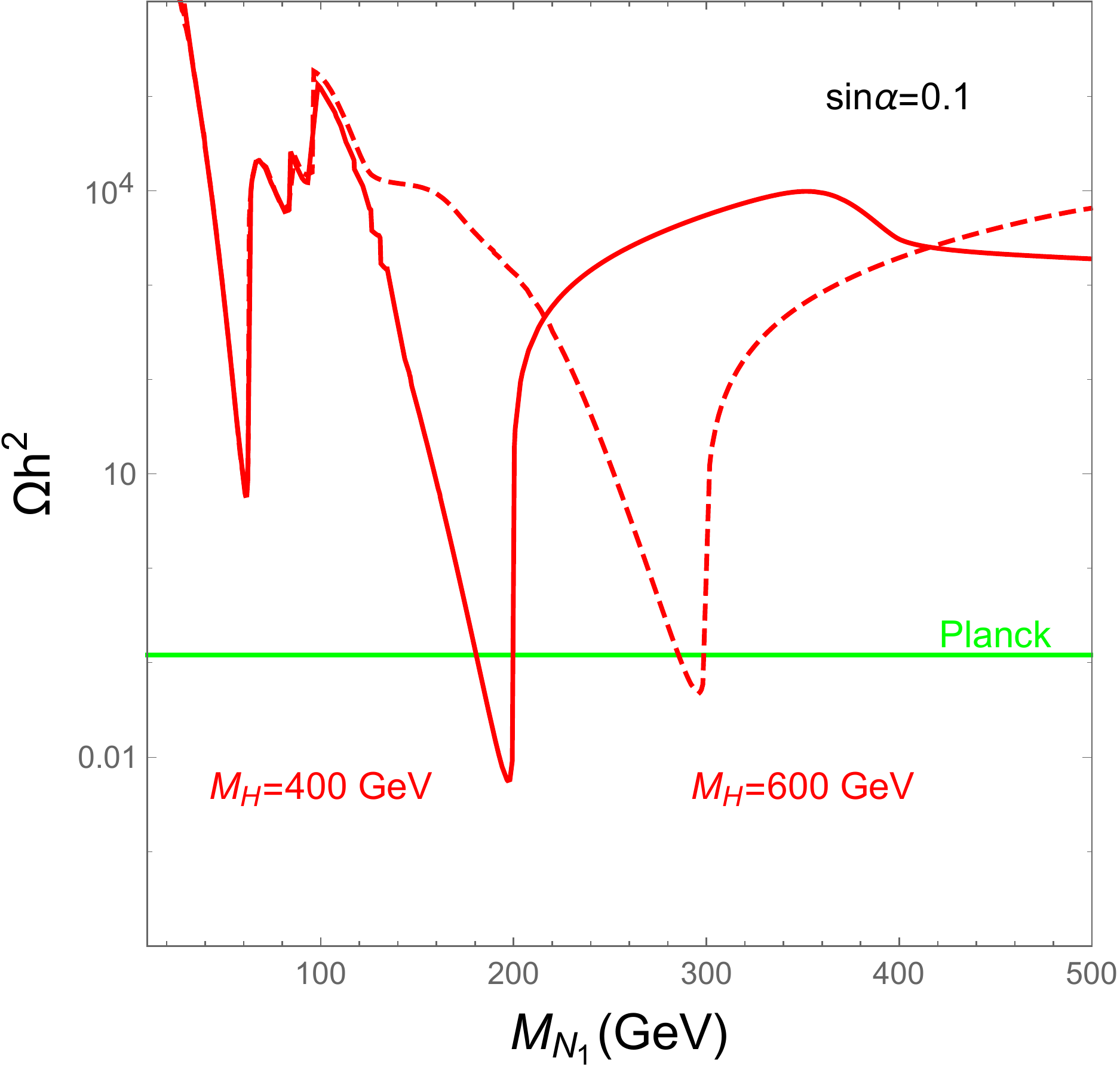}
\includegraphics[width=0.45\linewidth]{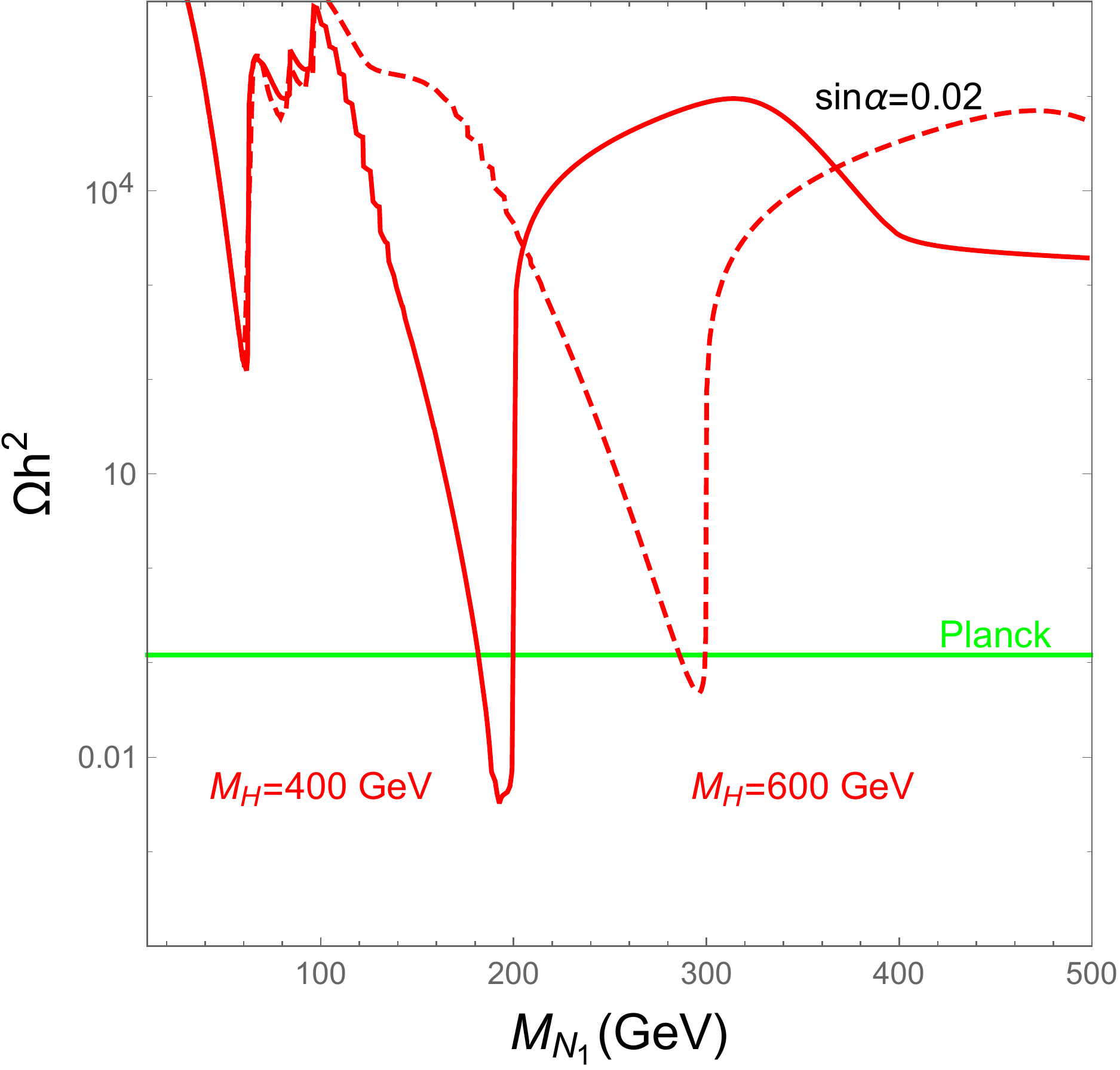}
\end{center}
\caption{Dark matter relic density as a function of $M_{N_1}$ for $\sin\alpha=0.1$ (left) and $\sin\alpha=0.02$ (right). The green line denotes Planck result $\Omega h^2=0.120\pm0.001$ \cite{Aghanim:2018eyx}.
 \label{Fig:DM}}
\end{figure}

In this paper, we consider the scalar singlet $H$ at EW scale and $Z'$ at TeV scale. Hence, the contributions to DM relic density from $Z'$-portal are suppressed by the heavy $Z'$ mass.  The {\tt micrOMEGAs} package \cite{Belanger:2018mqt} is employed for the calculation of DM relic density and DM-nucleon scattering cross section. The numerical results for relic density are depicted in Fig.~\ref{Fig:DM}. It is clear that the SM Higgs-portal can not lead to correct relic density for both $\sin\alpha=0.1$ and $\sin\alpha=0.02$. While resonance condition $M_{N_1}\sim M_{H}/2$ is always required to generate correct relic density for $H$-portal.

 \begin{figure}
\begin{center}
\includegraphics[width=0.45\linewidth]{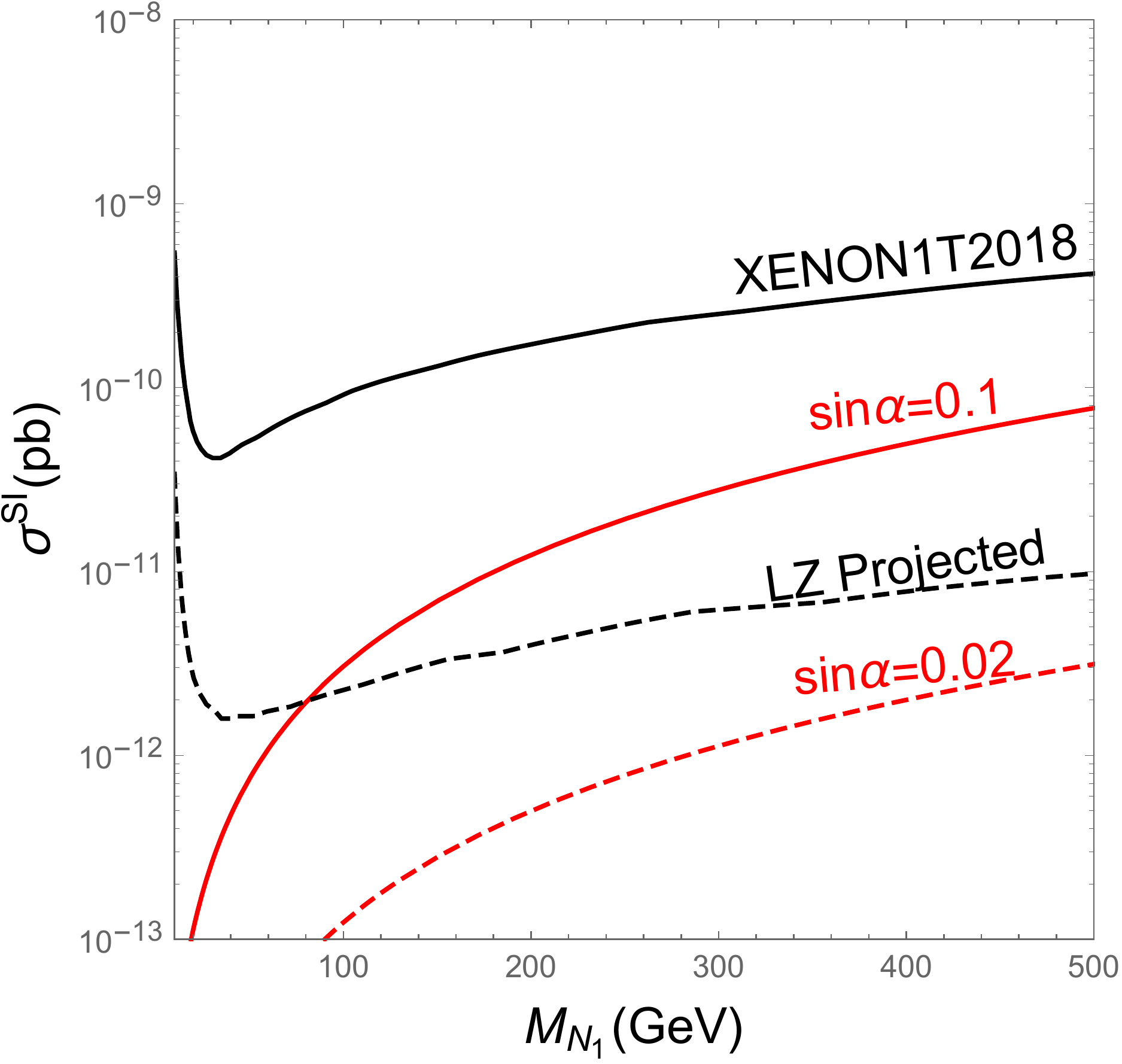}
\includegraphics[width=0.44\linewidth]{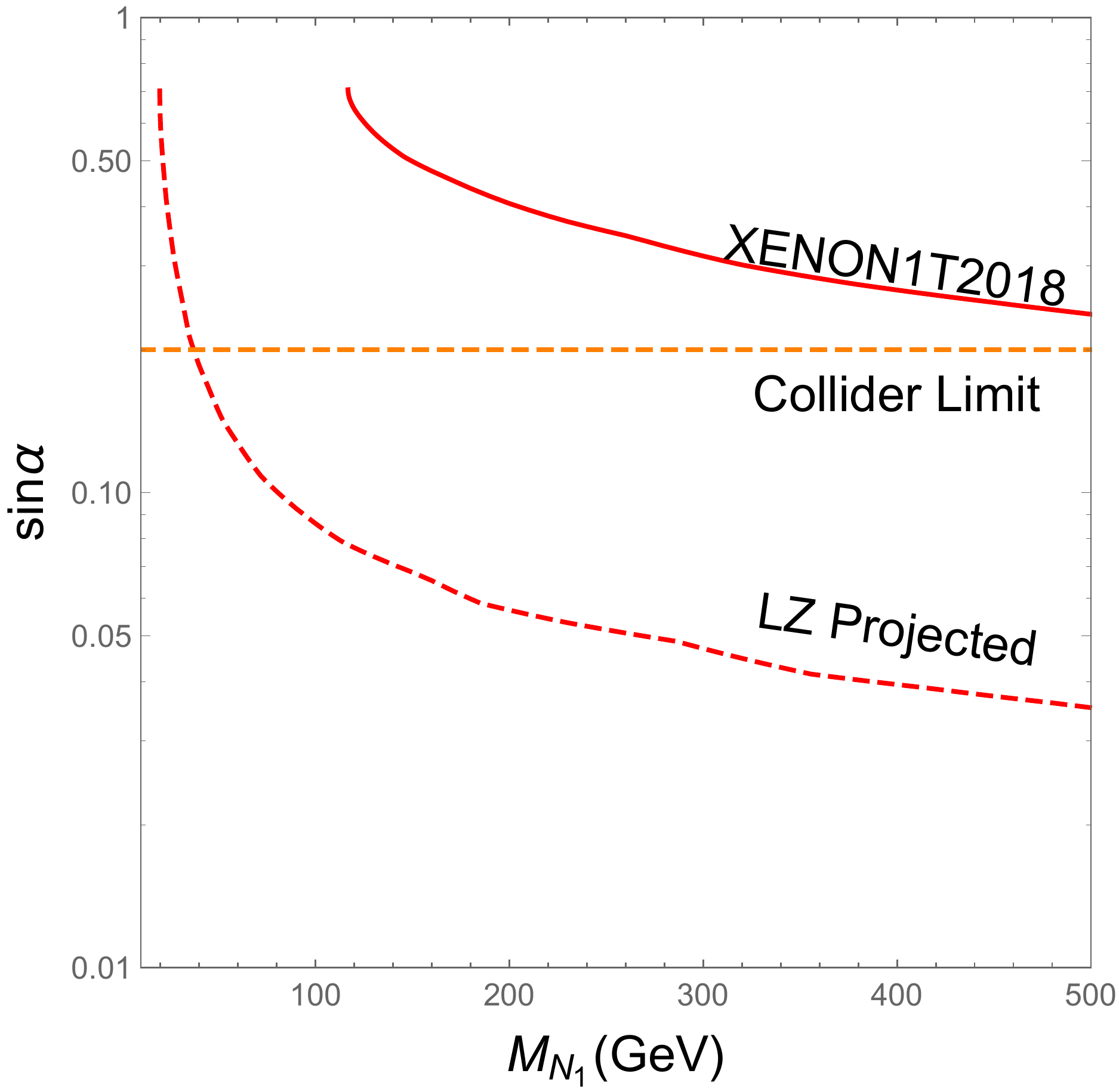}
\end{center}
\caption{Left: Spin-independent cross section as a function of $M_{N_1}$. The black solid and dashed line correspond to current XENON1T \cite{Aprile:2018dbl} and future LZ \cite{Akerib:2018lyp} limits, respectively. Right: Exclusion limits of XENON1T and LZ in the $M_{N_1}-\sin\alpha$ plane.
 \label{Fig:DD}}
\end{figure}

The spin-independent DM-nucleon scattering cross section is calculated as
\begin{equation}
\sigma^\text{SI}=\frac{4}{\pi}\left(\frac{M_p M_{N_1}}{M_p+M_{N_1}}\right)^2 f_p^2,
\end{equation}
where $M_p$ is the proton mass and the hadronic matrix element $f_p$ is
\begin{equation}
\frac{f_p}{M_p}=\sum_{q=u,d,s}f^p_{Tq} \frac{\alpha_q}{M_q}+
\frac{2}{27}\left(1-\sum_{q=u,d,s}f^p_{Tq}\right)\sum_{q=c,b,t}\frac{\alpha_q}{M_q}.
\end{equation}
with the effective vertex
\begin{equation}
\frac{\alpha_q}{M_q}=-\frac{y_{N_1}}{\sqrt{2}v}\sin2\alpha
\left(\frac{1}{M_h^2}-\frac{1}{M_{H_1}^2}\right).
\end{equation}
The parameters $f^p_{Tq}$ can be found in Ref.~\cite{Ellis:2000ds}.
Fig.~\ref{Fig:DD} shows the results for DM-nucleon scattering. It is clear that the choice of $\sin\alpha=0.1$ is enough to escape current XENON1T limit, but most range is in the reach of LZ. Since $\sigma^\text{SI}$ is sensitive to $\sin\alpha$,  we further derive the corresponding experimental limits. We find current XENON1T limit on $\sin\alpha$ is actually less stringent than collider limit in Ref.~\cite{Robens:2016xkb}. Meanwhile, if no signal is observed at LZ, then $\sin\alpha\lesssim0.05$ should be satisfied.

\subsection{Collider Signature}
\begin{figure}
\begin{center}
\includegraphics[width=0.45\linewidth]{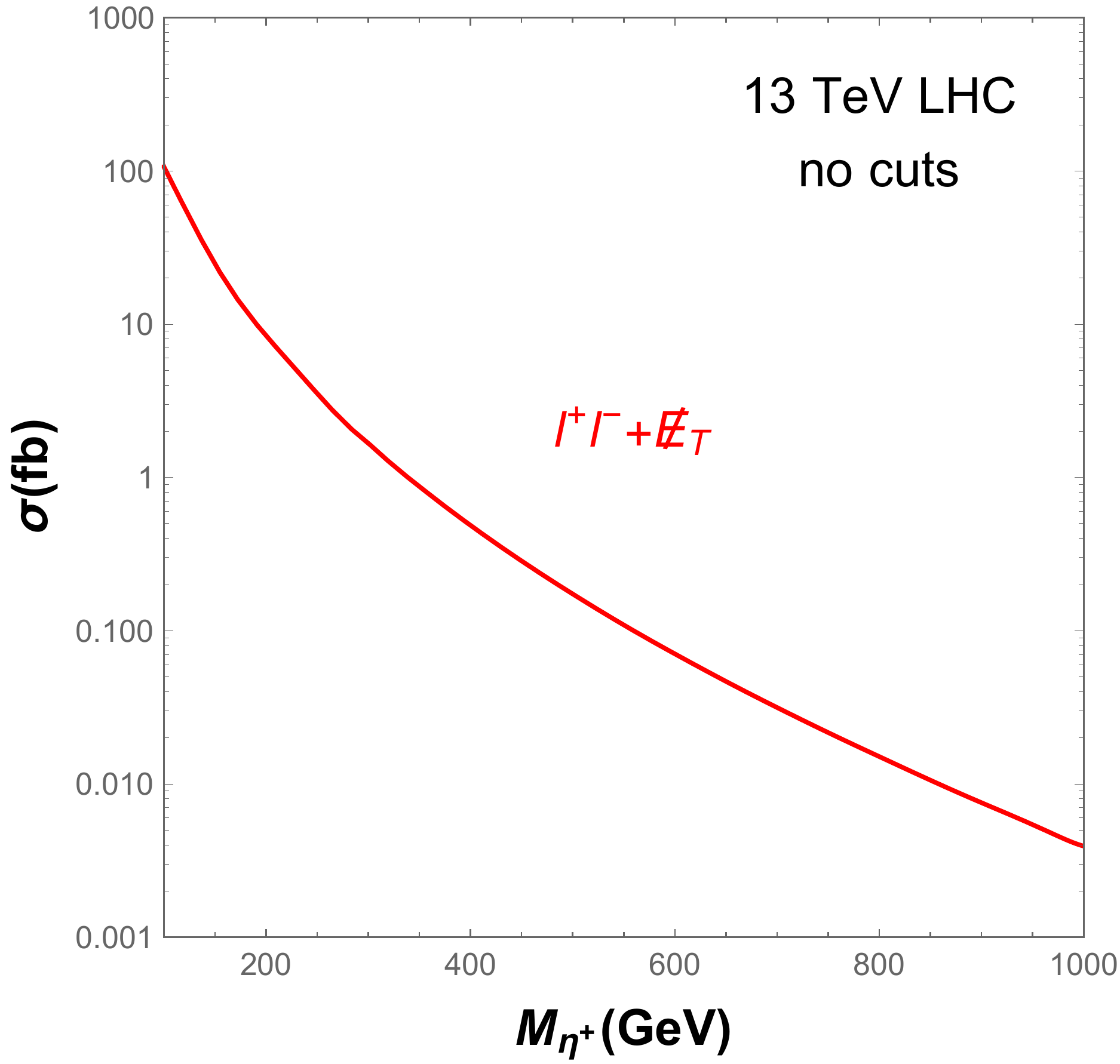}
\includegraphics[width=0.45\linewidth]{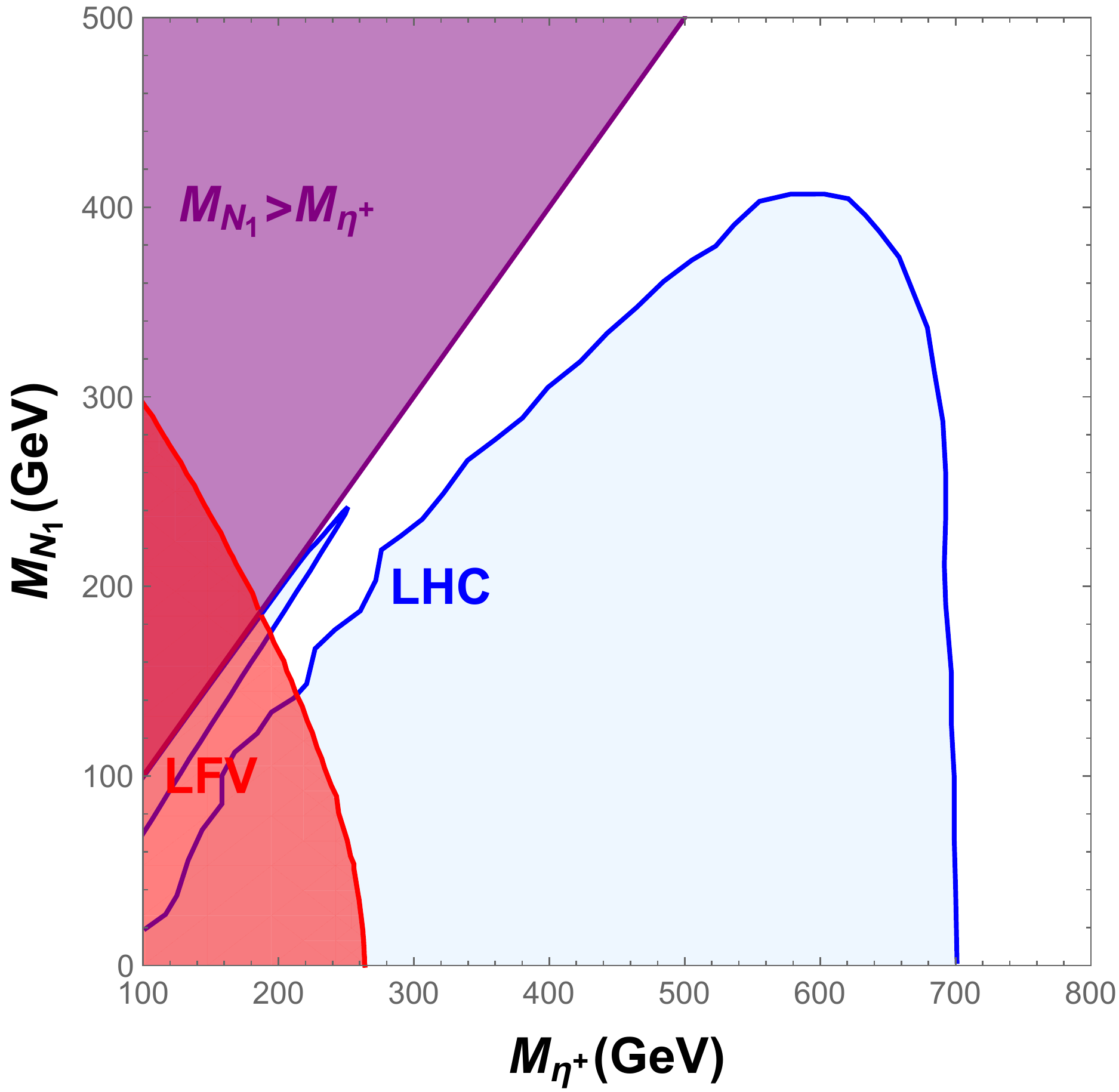}
\end{center}
\caption{Left: Theoretical cross section of $pp\to \eta^+\eta^-\to \ell^+\ell^-+\cancel{E}_T$ signature at 13 TeV LHC. Right: Excluded region in the $M_{N_1}$-$M_\eta$ plane by LHC direct search. The current LFV exclusion region is obtained by assuming $h'=0.01$.
 \label{Fig:DL}}
\end{figure}
With all new particles around TeV scale, various promising signatures can be probe by LHC \cite{Han:2019lux}. For simplicity, we consider a degenerate mass spectrum of the inert Higgs doublet $\eta$, i.e., $M_{\eta^\pm}=M_{\eta_R}=M_{\eta_I}$.
One promising signature is originated from the pair production of $\eta^\pm$ at LHC, i.e., $pp\to \eta^+\eta^-\to \ell^+N_1+\ell^-N_1$. Since $N_1$ is the DM candidate, this signature is just $\ell^+\ell^-+\cancel{E}_T$ \footnote{ The LHC limits are obtained with $\ell=e,\mu$, while there would be $\tau$ leptons in our scenario. Here, we directly take such limits as a conservative estimation, because for non-universal Yukawa structure, the $\tau$ final states can be suppressed under certain circumstance.}.  In the left panel of Fig.~\ref{Fig:DL}, theoretical production cross section at 13 TeV is shown, which is calculated by the help of {\tt MadGraph5\_aMC@NLO} \cite{Alwall:2014hca}. Meanwhile, exclusion region by LHC direct searches are shown in the right panel \cite{Aaboud:2018jiw,Aad:2019vnb,Aad:2019qnd}. Two viable regions are now allowed by direct search.  One is the heavy scalar region $M_\eta>700$ GeV and the other is the degenerate region $150~\GeV\lesssim M_{N_1}\lesssim M_{\eta}<700$ GeV. Together with the parameter space in Fig.~\ref{Fig:PS}, we find that the heavy scalar region might be probe by both LHC and $\mu-e$ conversion when $M_{N_1}\lesssim 200$ GeV. For $M_{N_1}\gtrsim200$ GeV or $M_{\eta^+}\gtrsim 800$ GeV, only LHC has the ability to probe. Meanwhile future upgrade $\mu\to 3e$ and  LHC searches with compressed mass spectra are both hopeful to probe the degenerate region. Note that the LFV exclusion limits depend on the choice of Yukawa coupling $h'$, i.e., $h'=0.01$, but the LHC searches do not.

Another promising signature is the dilepton signature from $pp\to Z'\to \ell^+\ell^-$. Neglecting final states mass, the corresponding partial decay widths are given by
\begin{eqnarray}
\Gamma(Z'\to f\bar{f})&=&\frac{M_{Z'}}{24\pi}g'^2 N_C^f (Q_{fL}^2+Q_{fR}^2),\\
\Gamma(Z'\to SS^*)&=&\frac{M_{Z'}}{48\pi} g'^2 Q_{S}^2,
\end{eqnarray}
where $N_C^f$ is the number of colours of the fermion $f$, i.e., $N_C^{l,\nu}=1$, $N_C^{q}=3$, and $Q_X$ is the $U(1)_{B-3L_\mu}$ charge of particle $X$. The branching ratio of $Z'$ are presented in Tab.~\ref{Tab:Zp}, which is clear that the dominant decay channel is $Z'\to \mu^+\mu^-$. The $B-3L_\mu$ nature of $Z'$ can be confirmed by determining the branching ratios as
\begin{equation}
\text{BR}(Z'\to b\bar{b}):\text{BR}(Z'\to e^+e^-):\text{BR}(Z'\to \mu^+\mu^-):\text{BR}(Z'\to \tau^+\tau^-)=
\frac{1}{3}:0:9:0.
\end{equation}
Such intrinsic property is useful to distinguish the $U(1)_{B-3L_\mu}$ gauge boson from other kinds of gauge bosons \cite{Basso:2008iv,Chun:2018ibr}.

\begin{table}
\begin{tabular}{|c|c|c|c|c|c|c|}
\hline
$q\bar{q}$ & $e^+e^-$  & $\mu^+\mu^-$ & $\tau^+\tau^-$& $\nu\nu$ & $NN$ & $H H$
\\
\hline
~0.090~ & ~0~ & ~0.405~ & ~0~ & 0.202 & 0.202 & 0.101
  \\ \hline
\end{tabular}
\caption{Decay branching ratio of $U(1)_{B-3L_\mu}$ gauge boson $Z'$, where we have show the lepton flavor individually.}
\label{Tab:Zp}
\end{table}

\begin{figure}
\begin{center}
\includegraphics[width=0.45\linewidth]{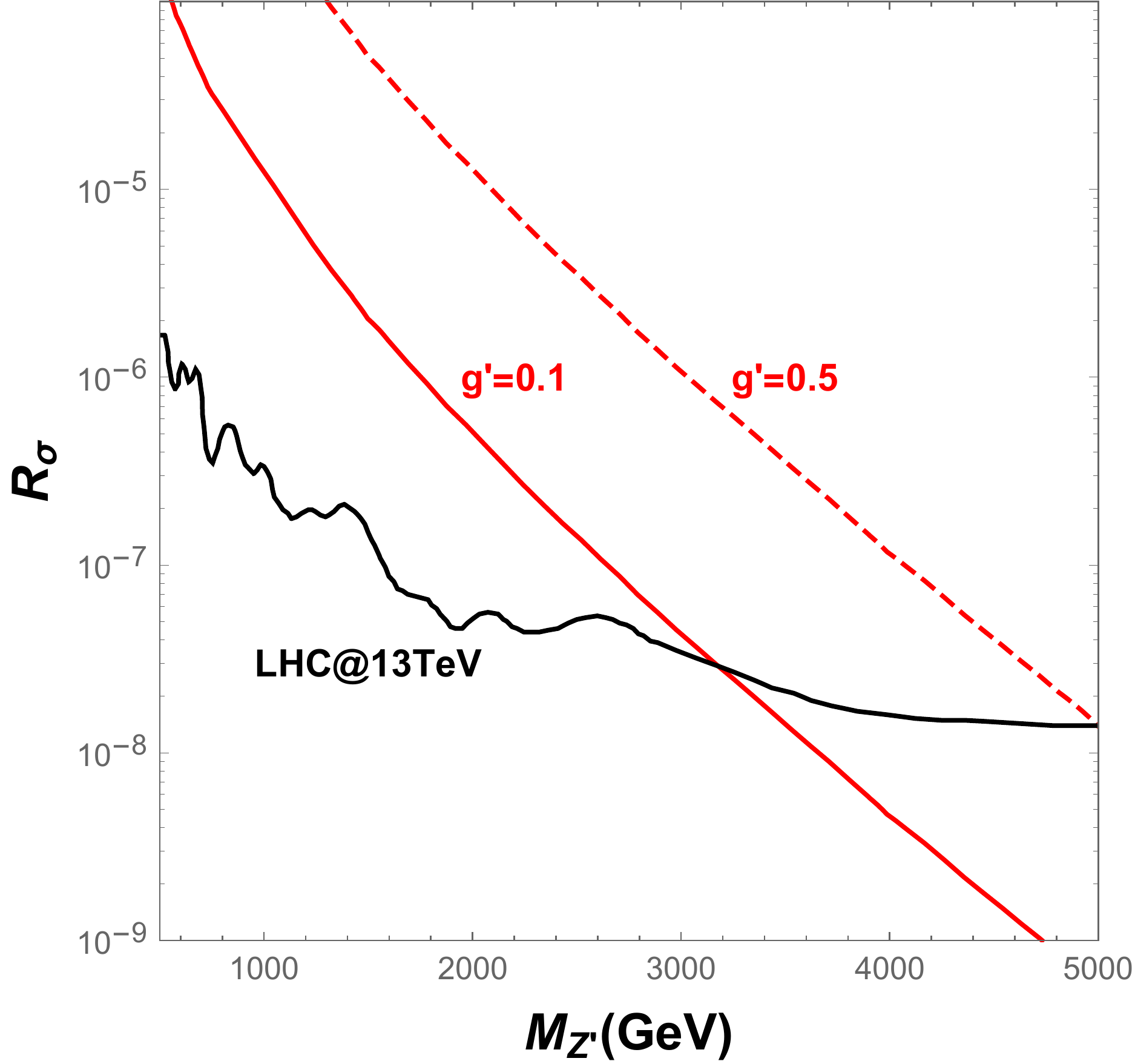}
\includegraphics[width=0.445\linewidth]{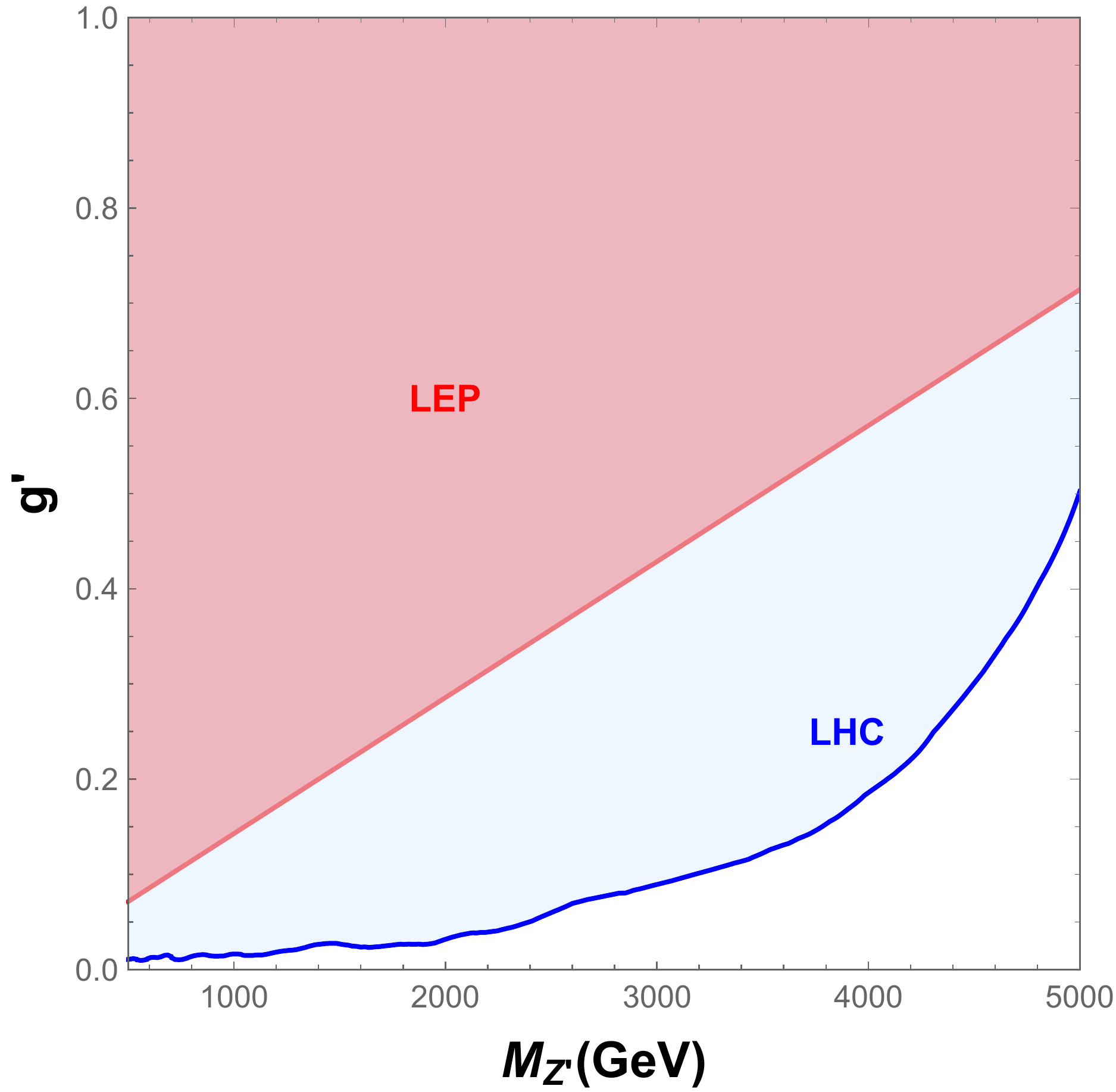}
\end{center}
\caption{ Left: Predicted cross section ratios in $U(1)_{B-3L_\mu}$ and corresponding limit from LHC. Right: Allowed parameter space in the $g'$-$M_{Z'}$ plane.
\label{Fig:Zp}}
\end{figure}
ATLAS \cite{Aaboud:2017buh,Aad:2019fac} and CMS \cite{Sirunyan:2018exx,CMS:2019tbu} have already perform searches for the $Z'$ by the dilepton signature. Since $Z'\to \mu^+\mu^-$ is the only dilepton final states, the CMS limit on $\mu^+\mu^-$ is more suitable for this model. The limit is on the ratio
\begin{equation}
R_\sigma=\frac{\sigma(pp\to Z'+X\to \mu^+\mu^-+X)}{\sigma(pp\to Z+X\to \mu^+\mu^-+X)}.
\end{equation}
Theoretical value and experimental limits are shown in left panel of Fig.~\ref{Fig:Zp}, which indicates that $M_{Z'}\gtrsim3.0(4.4)$ TeV should be satisfied for $g'=0.1(0.5)$. Exclusion region in the $g'$-$M_{Z'}$ plane is shown in the right panel of Fig.~\ref{Fig:Zp}. The LEP limit is now weaker than LHC limit for $M_{Z'}\lesssim5$ TeV.

\subsection{Combined Analysis}

Based on the discussion of previous benchmark scenario, we now perform a scan to search for combined allowed parameter space by neutrino data, lepton flavor violation, dark matter, and collider signature. For simplicity, we consider the simplified scenario discussed in Eq.~\eqref{eq:TZs} further with the assumption $h_{\alpha i}=h'$. In this way, $M_\nu$ and $M_N$ have exactly the same structure, and are related by $M_\nu = -\frac{\lambda v^2 h'^2}{32 \pi M_\eta^2} M_N$. Using neutrino oscillation data, one can obtain $M_\nu$, and then $M_N$ directly. Note for scenario $B-3L_\mu$, only the IH is allowed.
Besides IH neutrino oscillation data discussed in Sec.~III, we scan the other relevant parameters in the following region
\begin{eqnarray}
h'\in[10^{-3},10^{-1}], \lambda\in[10^{-11},10^{-1}], M_{\eta} \in[M_{N_1},1000]~\GeV, \\ \notag
 \sin\alpha\in[0.01,0.2],M_H\in[20,1000]~\GeV,v_S\in[5,10]~\TeV.
\end{eqnarray}
During the scan, we require that the IH neutrino oscillation parameters interpret texture B, $m_{ee}<0.061$ eV, and dark matter relic density satisfies $\Omega h^2\in [0.117,0.123]$. Then we apply current limits from $\sum_\nu$, LFV, and LHC sequentially. From Fig.~\ref{Fig:DD}, it is clear that $\sin\alpha<0.2$ satisfy current direct detection limit, so we do not apply current XENON1T limit. At last, we apply the future limits from LFV and LZ.
\begin{figure}
\begin{center}
\includegraphics[width=0.46\linewidth]{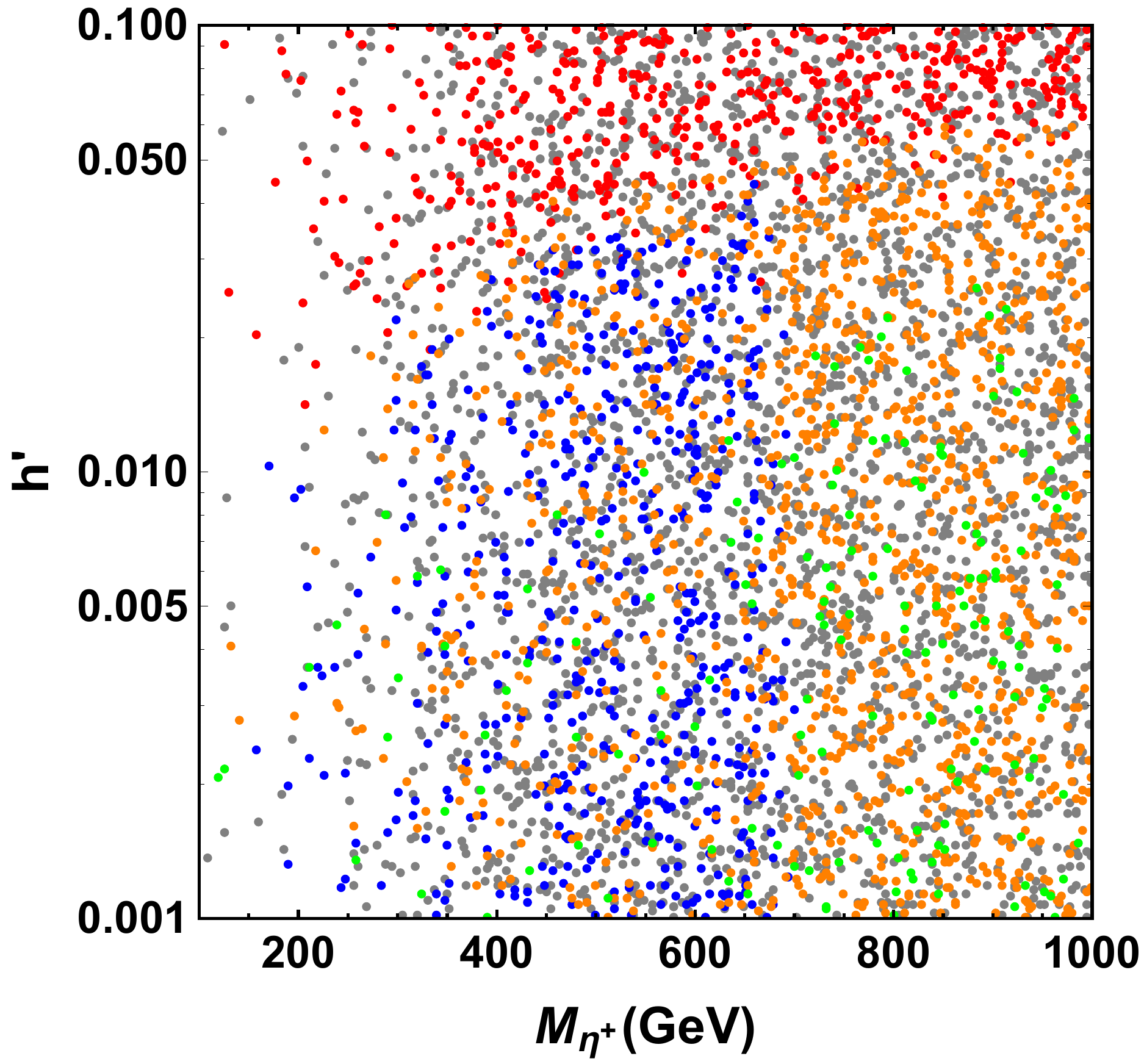}
\includegraphics[width=0.45\linewidth]{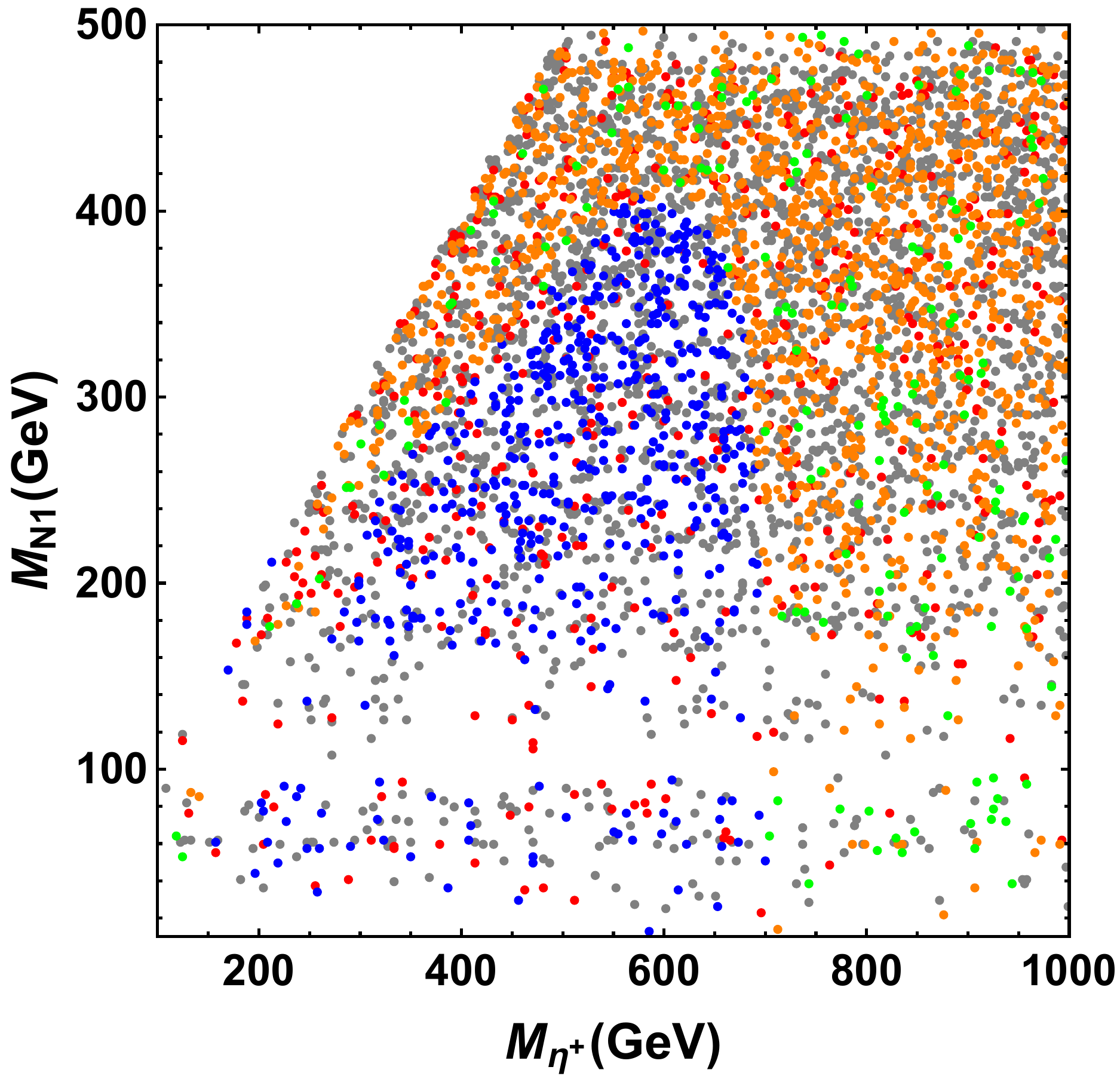}\\
\includegraphics[width=0.45\linewidth]{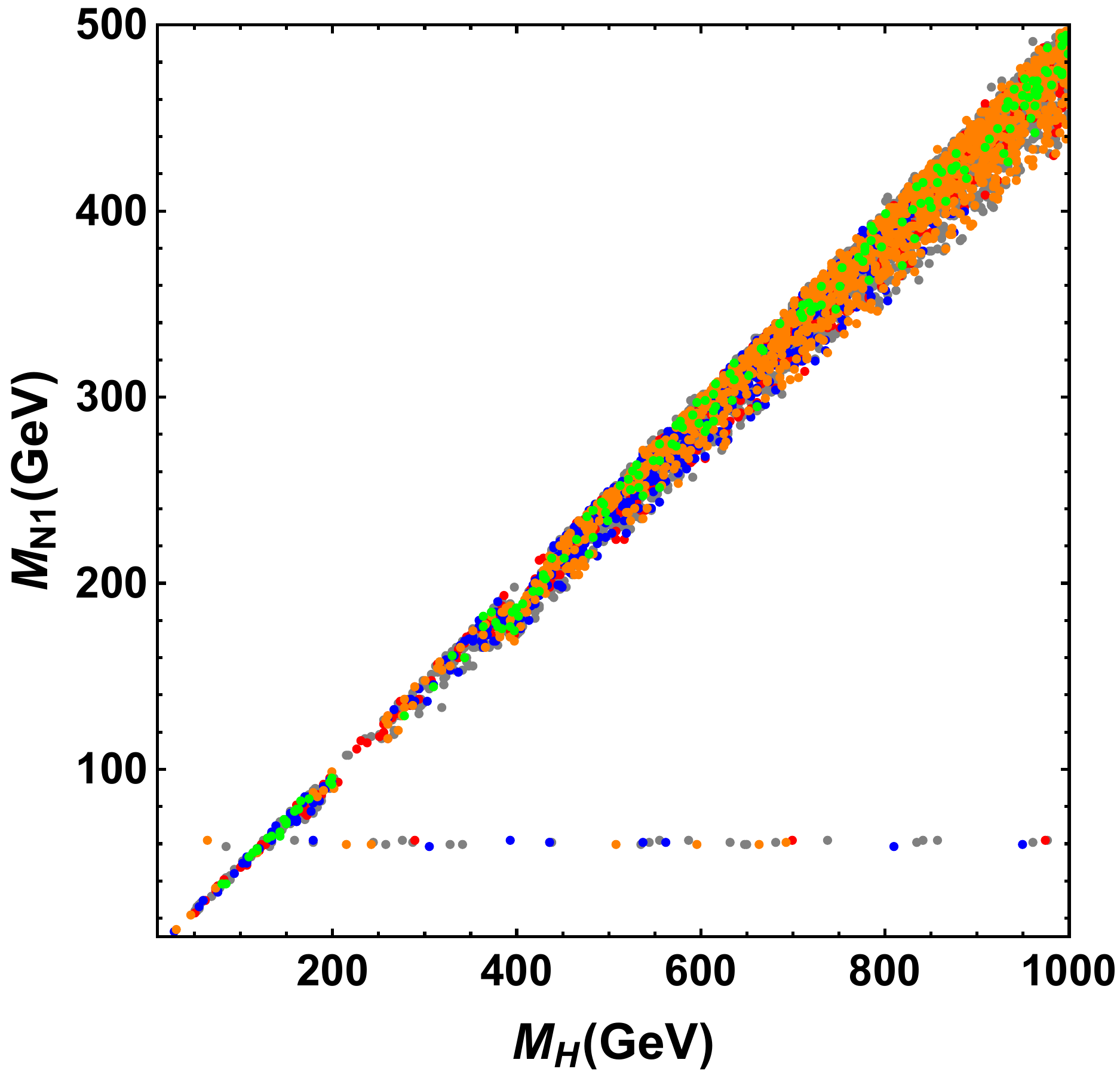}
\includegraphics[width=0.45\linewidth]{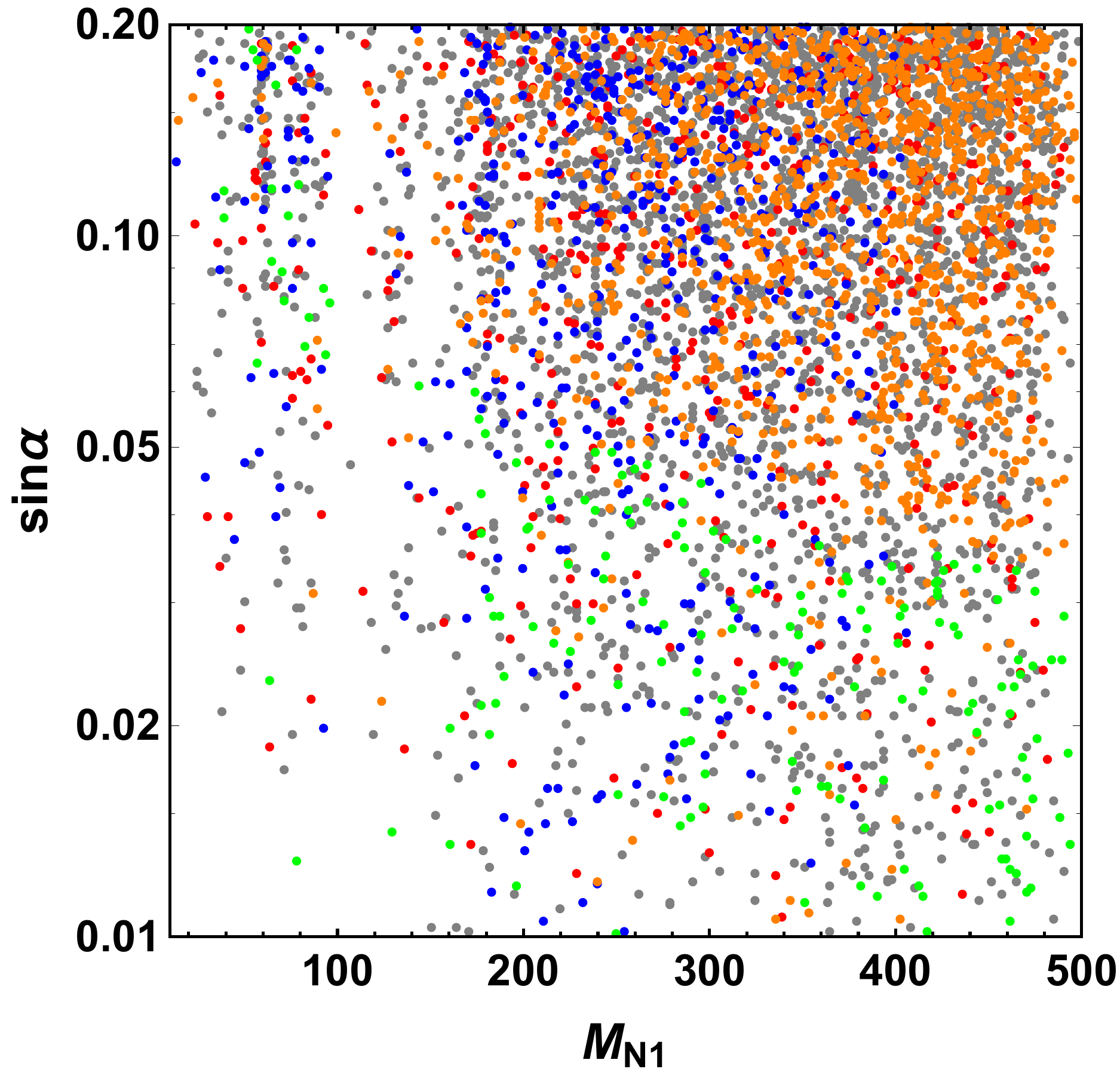}
\end{center}
\caption{ Scanning results for allowed parameter sets. The gray points are excluded by $\sum_\nu >0.12$ eV. The red points are excluded by LFV. The blue points are excluded by LHC. The orange points are within future reach of LFV and LZ. The green points are beyond future sensitivity.
\label{Fig:Scan}}
\end{figure}
The corresponding excluded and survived points are shown in Fig.~\ref{Fig:Scan}. Although the $\sum_\nu<0.12$ eV limits is stringent, only the Dirac phase shown in Fig.~\ref{mmIH} has specific favored region, i.e., $\delta\lesssim100^\circ$ or $\delta\gtrsim250^\circ$. LFV processes are sensitive to Yukawa coupling $h'$, i.e., $h'\lesssim0.05$ at present for $M_{\eta^+}<1$ TeV. With such small Yukawa coupling, the DM $N_1$ dominantly annihilate via the scalar singlet $H$ to SM final sates. Thus, it requires $M_{N_1}\sim M_H/2$, which is clearly shown in the third panel of Fig.~\ref{Fig:Scan}. Note for large enough mixing, e.g., $\sin\alpha\sim0.2$, the SM Higgs $h$ portal is also possible. But the $h$ portal will be fully excluded by LZ. As for LHC, its limits are sensitive to the mass spectrum of $M_{N_1}$ and $M_{\eta^+}$ with our assumption. And the direct detection experiment is sensitive to $\sin\alpha$. Finally, we show one benchmark point that satisfies all constraints
\begin{eqnarray}
\theta_{12}=35.59^\circ,\theta_{23}=46.52^\circ,\theta_{13}=8.325^\circ,
\delta=349.3^\circ,\rho=216.7^\circ,\sigma=292.1^\circ, \\\notag
m_1=0.05067~\eV,~m_2=0.05136~\eV,~m_3=0.009627~\eV,~\sum m_i=0.1117~\eV,\\\notag
h'=0.00365,M_{N_1}=239~\GeV, M_{\eta^+}=927~\GeV,M_H=483~\GeV,\sin\alpha=0.041.
\end{eqnarray}

\section{Conclusion}\label{Sec:CL}
\begin{center}
\begin{table}
\begin{tabular}{|c|c|c|c|c|c|c|}\hline\hline
Patten & Group & Hierarchy & Oscillation@$3\sigma$ & Oscillation@$1\sigma$
& $\sum_\nu<0.12$ eV & $M_{ee}$(eV)
\\ \hline
\multirow{2}*{A} & \multirow{2}*{$U(1)_{B-3L_e}$}
& NH & $\surd$ & $\surd$ & $\surd$ & 0
\\ \cline{3-7}
& & IH & $\times$ &  $\times$ &  $\times$ &  $\times$
\\ \hline
\multirow{2}*{B} & \multirow{2}*{$U(1)_{B-3L_\mu}$}
& NH & $\surd$ & $\times$ & $\times$ & $\gtrsim0.05$
\\ \cline{3-7}
& & IH & $\surd$ & $\surd$ & $\surd$ & $\gtrsim0.015$
\\ \hline
\multirow{2}*{C} & \multirow{2}*{$U(1)_{B-3L_\tau}$}
& NH & $\surd$ & $\surd$ & $\times$ & $\gtrsim0.033$
\\ \cline{3-7}
& & IH & $\surd$ & $\surd$ & $\surd$ & $\gtrsim0.015$
\\ \hline
\end{tabular}
\caption{Some main results of the one texture-zeros in the $U(1)_{B-3L_\alpha}$ scotogenic model.}
\label{TB:Res}
\end{table}
\end{center}
In this paper, we consider the flavor dependent $U(1)_{B-3L_\alpha}$ extension of scotogenic neutrino mass model. Within this framework, three kinds of one-zero-texture structures are realized in neutrino mass matrix. Predictions of the textures with latest neutrino oscillation parameters are performed. The main results are summarized in Table. \ref{TB:Res}. Therefore, we obtain three scenarios (patten A-NH, B-IH, C-IH)  favored by current experimental limits. Such three scenarios are distinguishable by the forthcoming experiments. Since future $0\nu\beta\beta$ experiments are hopefully to probe inverted hierarchy, patten A-NH is favored once no positive signature is observed. For patten B-IH, although oscillation parameters at $1\sigma$ range and $\sum_\nu<0.12$ eV are individually satisfied, they can hardly have common parameter space. So future precise measurements of neutrino oscillation data (especially the leptonic Dirac phase $\delta$) and sum of neutrino mass are able to exclude patten B-IH.
The corresponding gauge group can be confirmed by discovering a flavored  gauge boson $Z'$.

Then we have discussed phenomenologies such as lepton flavor violation, dark matter and collider signatures. For EW scale inert particles, the Yukawa coupling should be less than 0.01 to satisfy current tight constraints from $\mu\to e\gamma$. The DM candidate $N_1$ dominantly annihilates via the heavy scalar singlet $H$ with mass condition $M_H\sim2 M_{N_1}$. Observable direct detection signature is also mediate by $H$ for $\sin\alpha\gtrsim0.05$. The promising signature of inert particles at LHC is $pp\to \eta^+\eta^-\to \ell^+\ell^-+\cancel{E}_T$ with viable parameter space $M_{N_1}\lesssim M_\eta<700$ GeV or $M_\eta>700$ GeV. The $B-3L_\alpha$ nature of gauge boson $Z'$ can be confirmed by BR($Z'\to b\bar{b}$):BR($Z'\to \ell_\alpha\bar{\ell}_\alpha$)=$\frac{1}{3}:9$.

\section{Acknowledgements}

The work of Weijian Wang is supported by National Natural Science Foundation of China under Grant Numbers 11505062, Special Fund of Theoretical Physics under Grant Numbers 11447117 and Fundamental Research Funds for the Central Universities under Grant Numbers 2014ZD42. The work of Zhi-Long Han is supported by National Natural Science Foundation of China under Grant No. 11805081 , Natural Science Foundation of Shandong Province under Grant No. ZR2019QA021, No. ZR2018MA047.


\end{document}